\documentclass[12pt]{iopart}
\usepackage{graphicx,epsf}

\def\tree{{\rm tree}}
\def\pol{\varepsilon}

\newbox\charbox
\newbox\slabox
\def\slash#1{{      
        \setbox\charbox=\hbox{$#1$}
        \setbox\slabox=\hbox{$/$}
        \dimen\charbox=\ht\slabox
        \advance\dimen\charbox by -\dp\slabox
        \advance\dimen\charbox by -\ht\charbox
        \advance\dimen\charbox by \dp\charbox
        \divide\dimen\charbox by 2
        \raise-\dimen\charbox\hbox to \wd\charbox{\hss/\hss}
        \llap{$#1$} }}

\def\ksl{\slash{k}}
\def\qsl{\slash{q}}
\def\s{\sigma}
\def\spa#1.#2{\left\langle#1\,#2\right\rangle}
\def\spb#1.#2{\left[#1\,#2\right]}
\def\Tr{\, {\rm Tr}}

\def\NeqEight{{{\cal N}=8}}
\def\NeqFour{{{\cal N}=4}}
\def\NeqTwo{{{\cal N}=2}}
\def\NeqOne{{{\cal N}=1}}

\def\be{\begin{equation}}
\def\ee{\end{equation}}
\def\bea{\begin{eqnarray}}
\def\eea{\end{eqnarray}}
\def\ba{\begin{eqnarray}}
\def\ea{\end{eqnarray}}

\def\sect#1{section~{\ref{#1}}}
\def\eqn#1{eq.~(\ref{#1})}
\def\Eqn#1{Equation~(\ref{#1})}
\def\eqns#1#2{eqs.~(\ref{#1}) and~(\ref{#2})}

\def\fig#1{fig.~{\ref{#1}}}
\def\Fig#1{Figure~{\ref{#1}}}

\def\tree{{\rm tree}}
\def\oneloop{{\rm 1\hbox{-}loop}}

\def\Lloop{{L\rm\hbox{-}loop}}

\def\eps{\epsilon}
\def\e{\epsilon}

\def\tf{\tilde{f}}
\def\ib{{\bar\imath}}
\def\jb{{\bar\jmath}}

\newif\ifdraft
\drafttrue
\newif\ifpreprint
\preprinttrue

\begin{document}
\rightline{SLAC--PUB--14447}
\rightline{CERN--PH--TH/2011-092}

\title[Introduction to scattering amplitudes]{Scattering amplitudes: the most
perfect microscopic structures in the universe}

\author{Lance J. Dixon}

\address{Theory Group, Physics Department,
CERN, CH--1211 Geneva 23, Switzerland\\ 
and\\
SLAC National Accelerator Laboratory,
Stanford University,
Stanford, CA 94309, USA}
\ead{lance@slac.stanford.edu}
\begin{abstract}
This article gives an overview of many of the recent developments in
understanding the structure of relativistic scattering amplitudes in gauge
theories ranging from QCD to $\NeqFour$ super-Yang-Mills theory, as
well as (super)gravity.  I also provide a pedagogical introduction to some
of the basic tools used to organize and illuminate
the color and kinematic structure of amplitudes.
This article is an invited review introducing a special issue of 
Journal of Physics A devoted to ``Scattering Amplitudes in Gauge Theories''.
\end{abstract}

\maketitle

\section{Overview}
\label{IntroSection}

The astrophysicist Subrahmanyan Chandrasekhar once referred to black
holes as ``the most perfect macroscopic objects there are in the
universe: the only elements in their construction are our concepts of
space and time.''\,\cite{Chandrasekhar} This special issue of Journal
of Physics A is devoted to a completely different topic, the
quantum-mechanical amplitudes for the scattering of relativistic
particles, in which a revolution in our understanding has been taking
place recently.  However, scattering amplitudes represent a kind of
short-distance complement to the perfection Chandrasekhar saw in
black holes.  We might consider them to be the most perfect
microscopic structures in the universe.

Black holes are gigantic and (almost) eternal.  Scattering amplitudes,
in contrast, portray events that wink in and out of existence much faster than
the blink of an eye. They describe processes at the shortest distance scales
that can be probed in the laboratory.  When gravitons scatter, interacting
via Einstein's field equations, again
in some sense the only elements in the construction are our concepts
of space and time --- although the boundary conditions at infinity can
be more detailed than for black holes, involving multiple plane-wave
ripples (the gravitons) propagating in various directions through
Minkowski space-time.  When other relativistic particles scatter,
as in non-abelian gauge theory, more information has to be provided,
such as the gauge group, the coupling constant,
the matter content of the theory, the particle masses,
and the precise set of particle types and spins being scattered.
However, in the ultra-relativistic limit, if there are only gauge
interactions, much of the group-theoretical information can
be stripped from the amplitudes, and certain color-ordered
``primitive'' amplitudes can be defined which are quite universal.
These amplitudes have intricate analytic properties, which have
been teased out over the decades, but especially in the past few years.
A variety of new, efficient techniques have been found to construct
amplitudes, and hidden symmetries have been identified which illustrate
the amplitudes' ``perfection.''

In the real universe, black holes are not found in an absolute
vacuum.  Typically they are surrounded by accretion disks of gas and dust,
which muddy their perfection somewhat, but also enrich the range of
phenomena involving them, and make them much easier to observe,
albeit indirectly.  Similarly, scattering amplitudes
for quantum chromodynamics, or QCD, the most relativistic non-abelian
gauge theory we can study in the laboratory,
are shrouded from direct view by confinement.  The quarks and gluons that
theorists scatter mentally are never seen experimentally.  In the initial
state of a hadronic collision they are bound into a proton or other hadron.
In the final state they emerge as collimated jets of particles.
Despite this fact, quantities that are sensitive only to very short distances,
so-called infrared-safe observables such as jet production cross sections, 
can be computed reliably in terms of quarks and gluons,
in a systematic expansion in the strong coupling $\alpha_s$.

If we could go to asymptotically high energies,
so that $\alpha_s$ were infinitesimal, it would be enough to consider just
the leading order in this expansion, which is generated by the Born
approximation, or tree-level amplitudes.  The tree-level amplitudes
of QCD have even more ``perfection'' than the generic, loop-level amplitudes,
because they coincide with the tree amplitudes of a much more symmetric
theory, $\NeqFour$ supersymmetric Yang-Mills theory ($\NeqFour$ sYM).
This theory, which is the subject of many of the articles in this special
issue, has the maximum amount of supersymmetry possible in a gauge
theory.  Its four supercharges can be used to transform the helicity
$+1$ gluon state, by $\frac12$ unit of helicity at a time,
all the way to the helicity $-1$ CPT conjugate gluon state.
All states in the theory belong to a single supermultiplet,
transforming in the adjoint representation of the gauge group.
Therefore all interactions are related by supersymmetry to the
triple-gluon vertex, and the theory has a single dimensionless coupling $g$.
The $\beta$ function for $\NeqFour$ sYM vanishes exactly for any 
gauge group, $\beta(g)=0$,
so that the theory is exactly conformally invariant at the quantum level,
as well as classically.  (The classical conformal invariance of QCD with
massless quarks is of course spoiled by its nonvanishing one-loop $\beta$
function, which leads to asymptotic freedom.)

In investigations of scattering amplitudes in $\NeqFour$ sYM,
the gauge group under consideration is usually SU$(N_c)$.
Quite often, the limit $N_c\to\infty$ is also taken,
at a fixed value of the 't~Hooft coupling, $\lambda = g^2 N_c$.
For $\lambda \ll 1$, perturbation theory can be applied,
and for large $N_c$ only planar Feynman diagrams contribute.
For $\lambda \gg 1$, perturbation theory breaks down, but
the AdS/CFT correspondence~\cite{Maldacena} can be used.
In the large-$N_c$ or planar limit of $\NeqFour$ sYM,
even more remarkable symmetries and relations emerge for scattering
amplitudes.  As we will discuss further below, these properties
include dual conformal (super)symmetry,
which is part of a larger Yangian invariance; a description of amplitudes
in terms of spaces of complex planes (Grassmannians);
and in terms of various types of twistors;
and a relation between scattering amplitudes and the expectation values
of Wilson loops for closed polygons bounded by light-like edges.
Many of these properties are intimately connected with a strong-coupling
picture of gluon scattering in terms of strings moving in anti-de-Sitter
space~\cite{AldayMaldacena}.

The exceptional simplicity and numerous hidden symmetries of $\NeqFour$ sYM
have made it a playground for theorists interested in scattering amplitudes
for a couple of decades.  This interest has accelerated rapidly
in the past few years.  Because the tree-level amplitudes of
QCD and $\NeqFour$ sYM coincide, the early discoveries about 
QCD helicity amplitudes, motivated by the physics of jet production,
were also discoveries about $\NeqFour$ sYM.  For example, the
Parke-Taylor formula~\cite{ParkeTaylor} for the maximal-helicity-violating
(MHV) sequence of $n$-gluon amplitudes, which is just a single-term
expression, even for an arbitrarily large value of $n$, is also consistent
with all the symmetries of $\NeqFour$ sYM~\cite{Nair}, including some that
were only unveiled decades later.

Many general properties of tree amplitudes were understood early on,
such as their factorization properties~\cite{ManganoParke} and the
($\NeqOne$ or $\NeqTwo$) supersymmetry Ward identities
that they obey~\cite{SWI,SWIQCD}.
However, a fuller understanding and exploitation of these properties have
come only in recent years.  For instance, the complete solutions of the
long-known $\NeqFour$ supersymmetry Ward identities, and the corresponding
$\NeqEight$ identities in supergravity, have been worked out recently 
by Elvang, Freedman and Kiermaier~\cite{EFK}, and are reviewed
here~\cite{EFKHere}.

Several years ago, Witten~\cite{WittenTwistor} discovered that the
twistor space developed decades earlier by Penrose~\cite{Penrose}
gave a natural description of tree-level amplitudes in massless
gauge theory.  As discussed in \sect{KinematicPreliminariesSection},
such amplitudes have a natural description in terms of two-component
Weyl spinors, a left-handed and a right-handed spinor for each external
state.  To go to twistor space, one performs a Fourier transform on
half the variables, namely the left-handed spinors,
leaving the right-handed ones alone.  Witten found that gauge theory
could be reformulated in terms of a topological string moving in twistor
space.   Gauge amplitudes turn out to be localized on particular
curves in twistor space.  In one approach, the curves are intersecting lines:
a single line for the simplest MHV amplitudes, a pair of lines for the 
next-to-maximally-helicity-violating (NMHV) amplitudes, and so on.
This version gave rise to the CSW or MHV rules for gauge
theory~\cite{CSW}, whose MHV vertices are a particular off-shell
continuation of the Parke-Taylor formulae.  These developments
are reviewed here by Brandhuber, Spence and Travaglini~\cite{BSTHere},
and by Adamo, Bullimore, Mason and Skinner~\cite{ABMSHere}.
The latter reference also discusses another class of twistors, namely
momentum twistors, which also provides a natural set of kinematic variables
for amplitudes, in that they automatically satisfy the kinematic
constraints of momentum conservation and the mass-shell conditions.
Finally, the correspondence between Wilson loops and scattering amplitudes
is described in this article~\cite{ABMSHere} from the perspective of
twistor space.

Britto, Cachazo, Feng and Witten (BCFW)~\cite{BCFW}
recognized that factorization properties are powerful enough to allow
the derivation of recursion relations~\cite{BCFTree} for tree amplitudes,
in terms of on-shell lower-point amplitudes, which are evaluated at particular
complex momenta.  This observation, reviewed here by Brandhuber, Spence and
Travaglini~\cite{BSTHere}, makes essential use of the analyticity,
or {\it plasticity}, of amplitudes; that is, how they vary under smooth
deformations of the kinematics.
BCFW embedded an amplitude $A=A(0)$ into a one-complex-parameter
family of on-shell amplitudes $A(z)$ with shifted complex momenta, and
associated the poles of $A(z)$ in the $z$ plane with factorization of
the amplitude onto simpler lower-point amplitudes.  One can also 
construct a supersymmetrized version of the BCFW recursion relation,
in which Grassmann variables $\eta^A$ associated with
$\NeqFour$ supersymmetry are shifted along with the
momenta~\cite{AHCKGravity,BHT2008,BSTHere}.
This recursion relation can be solved explicitly to yield all
tree amplitudes in $\NeqFour$ sYM~\cite{DrummondHenn}; the solution
and its many symmetries are reviewed here by Drummond~\cite{DrummondHere}.

Scattering amplitudes in massless gauge theories all have quite similar 
structure at tree level, once the color factors have been stripped
off, as discussed in \sect{ColorPreliminariesSection}.
On the other hand, the structure of loop amplitudes depends critically
on the theory.  The simplicity of one-loop multi-gluon amplitudes
in $\NeqFour$ sYM played a key role in the development of the unitarity
method~\cite{BDDKNeq4,BDDKNeq1}.  This method reconstructs loop amplitudes
from their unitarity cuts, again exploiting analyticity.
In general at one loop, as explained in this issue by Britto~\cite{BrittoHere}
and Ita~\cite{ItaHere},
unitarity cuts are matched against a decomposition of the
amplitude in terms of a set of scalar integrals --- boxes, triangles, bubbles
and (sometimes) tadpoles --- in order to determine the coefficients
of the integrals.  In $\NeqFour$ sYM, the high degree of supersymmetry
implies that only boxes have non-vanishing coefficients.  Using unitarity,
an infinite sequence of one-loop amplitudes (the MHV amplitudes)
could be determined in $\NeqFour$ sYM from just the product of two
tree-level MHV (Parke-Taylor) amplitudes~\cite{BDDKNeq4}.

Later, generalized unitarity was applied to the problem
of computing one-loop amplitudes, by extracting additional
information from products of three and four tree amplitudes,
the so-called triple~\cite{Z4partons} and quadruple~\cite{BCFUnitarity} cuts.
Quadruple cuts have four on-shell conditions.  These four equations
determine the 
four components of the (four-dimensional) loop momenta completely,
up to a (possible) two-fold degeneracy.  This realization allows the 
box coefficients to be computed very simply in terms of the product of
four tree amplitudes, glued together as the four corners of a box, and
evaluated at each of the two solutions for the loop
momenta~\cite{BCFUnitarity}.  For $\NeqFour$ sYM,
there are no other coefficients to determine, so the one-loop problem
has been ``reduced to quadrature.''  (It turns out that maximal
$\NeqEight$ supergravity can be proven to have the same ``no-triangle''
property~\cite{BjerrumVanhove,AHCKGravity}, so it too is solved
at one loop by the quadruple cuts.)

Generalized unitarity can also be applied very effectively
at the multi-loop level.  In principle it can be used for any gauge
(or gravitational) theory.  As an example, the two-loop four-gluon
scattering amplitudes in QCD have been computed in this way~\cite{TwoLoop4g}.
In practice the method has been pushed the furthest in $\NeqFour$ sYM.
The basic techniques of multi-loop generalized unitarity are reviewed
here by Bern and Huang~\cite{BernHuangHere},
and by Carrasco and Johansson~\cite{CJHere}.

Returning now to one-loop amplitudes for generic gauge theories,
including QCD, the triangle and bubble coefficients have to be determined,
as well as certain {\it rational parts}, which have no unitarity cuts in
four dimensions.
The triple cuts contain information about the triangle coefficients,
but they also receive contributions from boxes.  Similarly, the ordinary
double cuts determine the bubble coefficients, once the contributions
of triangles and boxes are removed.
This separation can be done analytically, by making use of the different
analytic behavior of the different types of cut
integrals, as reviewed by Britto~\cite{BrittoHere}.  It can also be done
numerically, by a suitable sampling of the continuum of loop momenta
solving the triple or double cut on-shell
conditions~\cite{OPP,EGK07,BlackHat08}, as described by
Ita~\cite{ItaHere}.  These two articles also review methods for
computing the rational parts of one-loop amplitudes.  Although
the rational parts have no unitarity cuts in four dimensions (by
definition), they can be constructed via their cuts in dimensional
regularization in $D=4-2\e$ dimensions.  Alternatively, it is possible
to construct them recursively in the number of legs~\cite{BlackHat08}.

On-shell loop amplitudes in massless non-abelian gauge theories always
contain infrared divergences, due to the exchange of soft gluons or
virtual collinear splittings.  In QCD it is conventional to regulate
these divergences, as well as the ultraviolet ones, using dimensional
regularization.  In supersymmetric theories, dimensional
reduction~\cite{DimRed}, or the related four-dimensional helicity
scheme~\cite{FDH}, can be used to keep the number of bosonic and
fermionic states the same, and preserve Ward identities for ordinary
supersymmetry.

\begin{figure}[t]
\centerline{\epsfxsize 4.5 truein \epsfbox{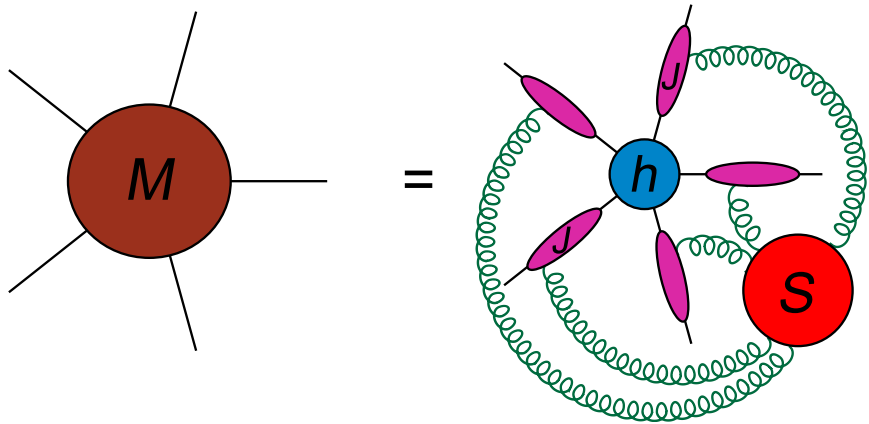}}
\caption[a]{\small Factorization of soft and collinear singularities.}
\label{SoftCollFactFigure}
\end{figure}

The general structure of the infrared divergences is well understood
from decades of work in QED as well as QCD~\cite{Sudakov}.  It has
been worked out in the most detail in the context of dimensional
regularization~\cite{MagneaSterman,CataniIR,TYS}.  The basic picture
is shown in \fig{SoftCollFactFigure}.  Soft divergences and collinear
divergences associated with the amplitude $M$ each have a universal
form.  They can be factorized from each other and from a hard,
short-distance part of the amplitude.  Soft divergences, denoted by
the blob marked $S$, come from exchange of long-wavelength gluons.
These gluons do not have the resolving power to probe the internal
structure of the ``jets'' $J$ of virtual collinear particles which
capture the collinear divergences.  Soft gluons can only see the
overall color charges of the jets.  An individual jet function $J_i$
depends on the type of particle $i$, but not on the full amplitude
kinematics.  The soft function $S$ does not depend on the particle
types, but only on their momenta and color quantum numbers.  In
general these quantum numbers can be mixed by gluon exchange, so
$S$ is a matrix in color space.  The hard function $h$ has no infrared
singularities, but generically depends on the particle types, colors
and kinematics.

\begin{figure}[t]
\centerline{\epsfxsize 4.5 truein \epsfbox{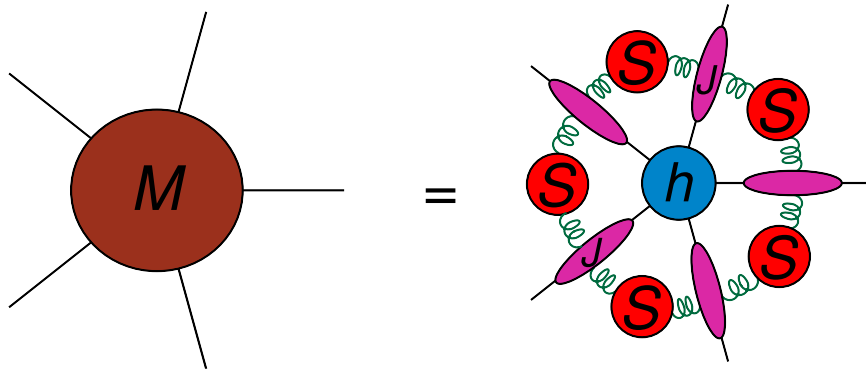}}
\caption[a]{\small Soft-collinear factorization in the planar limit.}
\label{SoftCollFactPlanarFigure}
\end{figure}

In the planar or large-$N_c$ limit, the picture simplifies
considerably, to that shown in \fig{SoftCollFactPlanarFigure}.  Now
$M$ represents the coefficient of a particular color structure, such
as $\Tr(T^{a_1}T^{a_2}\cdots T^{a_n})$ (assuming that all external
states are in the adjoint representation; see
\sect{ColorPreliminariesSection}).  In the planar limit, individual
soft gluons can only connect color-adjacent external partons.  There
is no mixing of different color structures at large $N_c$.  One can
absorb the entire soft function $S$ into jet functions, which
corresponds to breaking up the right-hand side of
\fig{SoftCollFactPlanarFigure} into $n$ wedges.  Each wedge is bounded
by two hard lines, and is composed of ``half'' of each of the two jet
functions, as well as the soft gluons exchanged between them.  Up to
nonsingular terms, the wedge controlling the infrared divergences
represents the square root of the Sudakov form factor, which is
defined as the amplitude for a color-singlet state to decay into a
pair of (adjoint) gluons.

In dimensional regularization, the Sudakov form factor obeys a
particular differ\-ential equation~\cite{MagneaSterman}, whose
source term is called the
cusp anomalous dimension $\gamma_K$~\cite{KM}.  There is an additional
constant of integration, called ${\cal G}_0$.  These two functions (along
with the $\beta$ function in a non-conformal theory) control
the infrared divergences to all orders for any amplitude
in a planar massless gauge theory.
In planar $\NeqFour$ sYM, the cusp anomalous dimension has been determined
to all orders in the coupling using integrability~\cite{BES}.
The BDS ansatz~\cite{BDS} was built up from this description of the infrared
singularities of planar amplitudes, plus the observation
for the two- and three-loop four-point amplitudes 
that the hard function $h$ essentially reduces to a constant,
independent of the kinematics.

Besides the cusp anomalous dimension, another place that integrability
certainly enters multi-loop scattering amplitudes is in
multi-Regge-kinematics, a particular class of high-energy or small-angle
limits of $2\to (n-2)$ particle scatterings.
These kinematics gave the first indication that the BDS
ansatz for MHV amplitudes
has to be corrected beginning at the six-point level~\cite{BLSV}.
(It was previously argued~\cite{AM2} using the properties of Wilson loops,
that the ansatz should be corrected at two loops for a large value of $n$, 
unless the correspondence between MHV amplitudes and Wilson loops were
to break down.  The hexagon Wilson loop was then computed and
found to differ from the ansatz prediction~\cite{HexagonWilson}.)
The dynamics of gluons in the transverse plane is given, in the
leading-logarithmic approximation, by the Hamiltonian for a integrable
open spin chain, as reviewed here by Bartels, Lipatov and
Prygarin~\cite{BLPHere}.

In planar $\NeqFour$ sYM, another infrared regulator is more
convenient for many purposes, in particular for exploring the
loop-level consequences of dual conformal symmetry, which is not
preserved by dimensional regularization.  The features of a
recently-developed ``Higgs regulator''~\cite{AHPS} are reviewed in
this issue by Henn~\cite{HennHere}.  For this regulator, vacuum
expectation values are given to some of the adjoint scalar fields in
the theory, breaking the gauge symmetry in such a way that (in the
planar limit) the amplitudes are regulated by massive particles
circulating around the outside of the loop diagrams.  Dual conformal
symmetry remains exact in a certain sense.  When the particle masses
$m_i$ are taken to be much less than the momentum-invariants for the
scattering process, logarithmic divergences develop, the analogs of
the $1/\e$ infrared poles in dimensional regularization.

\begin{figure}[t]
\centerline{\epsfxsize 5.2 truein \epsfbox{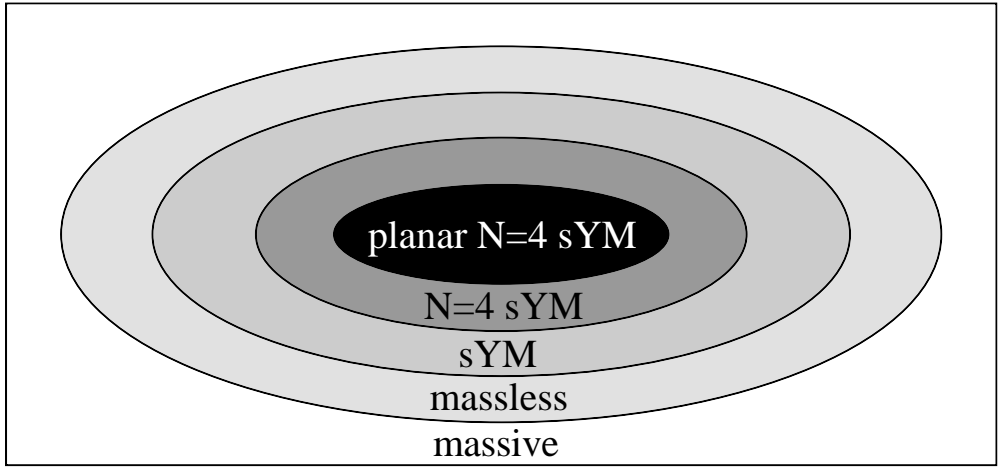}}
\caption[a]{\small Hierarchy of simplicity in scattering amplitudes
for various types of gauge theory.}
\label{GaugeTheoryFigure}
\end{figure}

There is a hierarchy of simplicity in the scattering amplitudes for
various types of gauge theory, as sketched in \fig{GaugeTheoryFigure}.
This hierarchy begins to be revealed at one loop.  The outer region of
the diagram stands for a generic gauge theory with massive matter
fields, and perhaps massive gauge bosons, if the gauge symmetry is
spontaneously broken, as in electroweak theory.  One-loop amplitudes
in such a theory generically contain tadpole integrals.  One-particle
cuts are nontrivial, and are particularly delicate because of
external-leg contributions~\cite{EGKMMass,BrittoHere}.  The cut
structure of loop integrals containing massive propagators in the loop
is generically somewhat more complicated than the purely massless
case.  Massive particles in the loop can be unstable, which usually
necessitates complex masses.  When one enters the ``massless'' ring in
\fig{GaugeTheoryFigure}, corresponding to massless gauge bosons and
matter fields, most of these complications vanish, although there are
still generically rational parts to compute.  The ring ``sYM'' stands
for supersymmetric gauge theories.  Their one-loop amplitudes can be
constructed from four-dimensional unitarity cuts alone, {\it
i.e.}~there are no non-trivial rational parts~\cite{BDDKNeq1}.

Moving further inward in \fig{GaugeTheoryFigure}, we arrive at
$\NeqFour$ sYM.  As mentioned earlier, at one loop the coefficients of
bubble and triangle integrals now vanish, as well as the independent
rational parts.  (There are other gauge theories with vanishing bubble and
triangle coefficients, at least for their $n$-gluon
amplitudes~\cite{LalRaju,BrittoHere}.)  The theory becomes conformally
invariant. It has been conjectured that the {\it leading
singularities} --- the multi-loop analogs of the quadruple cuts ---
are sufficient to determine the amplitudes at any loop
order~\cite{AHCKGravity}.  In addition, scattering amplitudes have
empirically a predictable, uniform transcendental
weight~\cite{KLOV,Neq44}.  This weight refers to their construction
out of polylogarithms, logarithms, and Riemann $\zeta(n)$ values.  For
example, the finite (${\cal O}(\e^0)$) terms in one-loop $\NeqFour$
sYM amplitudes are of weight two: They contain some terms proportional
to the polylogarithm ${\rm Li}_2$, and others which are products of
two logarithms, or proportional to $\zeta(2)$, but they do not contain
any terms of lower transcendentality.  (At one loop this result just
follows from the absence of bubble and tadpole integrals.)  Wherever
analytic results are available, this uniform transcendentality
property holds.  For example, at two loops (weight four for the finite
terms) it holds for the non-planar, subleading-color terms as well as
the planar ones~\cite{NNSSublNeq4}.

Finally, for planar $\NeqFour$ sYM, as mentioned earlier,
additional symmetries appear, dual superconformal invariance and an
associated Yangian symmetry. The extra symmetries are related to
integrability.  They give rise to the prospect of adapting methods
developed for other integrable systems to determine the $S$-matrix
exactly, at least in the large $N_c$ limit.
Indeed, as reviewed by Drummond~\cite{DrummondHere},
an integrand exhibiting manifest Yangian invariance 
has been proposed recently~\cite{NimaAllLoop}
for the general $L$-loop $n$-point amplitude,
based on the BCFW recursion relations;
see also ref.~\cite{BoelsAllLoop}.

At the level of integrated amplitudes, dual conformal
invariance has an anomaly due to the need for an infrared regulator.
The anomaly was first understood in terms of Wilson loops rather
than amplitudes~\cite{DHKSWLAnomaly} (for which divergences are
ultraviolet in nature, rather than infrared).
Nevertheless, dual conformal symmetry
completely fixes the form of the four- and five-point
amplitudes.  The form to which they are fixed turns out to be precisely
the ABDK/BDS ansatz~\cite{ABDK,BDS}, which was developed earlier
based on the structure of infrared divergences of amplitudes,
plus patterns observed at two and three loops.
As discussed in \sect{PlanarVariablesSubsection}, for six external
legs one can build three different combinations of kinematic variables
that are invariant under all dual conformal transformations,
the cross-ratios $u_1$, $u_2$ and $u_3$.
At the six-point level, the ABDK/BDS ansatz fails at the first
nontrivial order, two loops~\cite{SixPtFailure}, and
dual conformal symmetry allows for an additional
``remainder function'', $R_6^{(2)}(u_1,u_2,u_3)$, which
depends on the three cross-ratios.

At strong coupling, large $\lambda$ and large-$N_c$, Alday and
Maldacena used the AdS/CFT correspondence to map the problem of gluon
scattering to that of strings moving in anti-de-Sitter
space~\cite{AldayMaldacena}.  The AdS background is weakly curved in
this limit, relative to the string scale.  Another way of saying this
is that the two-dimensional string world-sheet stretches over a large
area, relative to the scale on which strings fluctuate, allowing a
semi-classical expansion to be used at large $\lambda$.  In the
leading term in the expansion, different helicity configurations are
not distinguished.  The amplitude is given simply by $\exp(-S_{\rm
cl})$, where $S_{\rm cl}$ is the classical action that minimizes the
area (at least for scattering configurations with a Euclidean
interpretation).  Using a $T$-duality transformation in string theory,
Alday and Maldacena showed that the boundary
conditions for the world-sheet are closed polygons, where each edge is
a light-like segment given by the momentum $k_i$ of the $i^{\rm th}$
gluon.  The solution to the minimal area problem was found explicitly
for four-gluon scattering.  More recently, in order to handle more complicated
kinematical configurations than the four-point case, the integrability 
of the string sigma-model action~\cite{BPR} has been exploited.
The minimal area problem has been solved by mapping it to a system
of equations identical to those of the Thermodynamical Bethe
Ansatz~\cite{AGMTBA}.

The first computation at strong coupling was for the four-gluon amplitude,
and matched precisely the prediction of the
ABDK/BDS ansatz~\cite{AldayMaldacena}.  But it also suggested
a weak-coupling correspondence between scattering amplitudes and
Wilson loops for closed light-like polygons, which was
rapidly established for a variety of cases, initially for the simplest
case of MHV scattering
amplitudes~\cite{MHVWilsonLoops,DHKSWLAnomaly,SixPtFailure}.
The Wilson-loop correspondence has been very important both conceptually
and technically.
Conceptually, it is still not entirely clear why it happens at all
at weak coupling.  Technically, given its existence, 
it allows amplitudes to be computed in terms of Wilson line integrals,
which are generally somewhat more tractable than Feynman loop integrals.
Some recent conceptual advances have come from considering
certain correlation functions with close-to-light-like
separations~\cite{RecentWilsonLoops,ABMSHere}.

The relative simplicity of the Wilson line integrals for the six-point
(hexagon) case allowed the remainder function $R_6^{(2)}(u_1,u_2,u_3)$
to be evaluated analytically in terms of Goncharov
polylogarithms~\cite{DDS}.  The expression in ref.~\cite{DDS} was then
simplified considerably, to a few lines involving classical
polylogarithms ${\rm Li}_m$, using properties of the symbol operation
which captures all of the essential analytic behavior of a
multi-variable function~\cite{Goncharov}.  These results open the door
to the possibility of finding simple analytic results for more general
kinematic configurations. For some configurations in which the external
momenta are restricted to two space-time dimensions, two-loop analytic
results are already available beyond six external legs~\cite{EightOrMore}.
There is certainly the prospect of more multi-loop analytic results 
on the horizon, for both Wilson loops and scattering amplitudes in
planar $\NeqFour$ super-Yang-Mills theory.

Gravity seems to be a completely separate force from gauge theory.  It
has a spin two force carrier instead of spin one, no color degrees of
freedom, and a dimensionful coupling constant.  Nevertheless, the two
theories are intimately connected.  The AdS/CFT
correspondence~\cite{Maldacena} relates the two, holographically, as a
{\it weak-strong} duality.  As mentioned earlier, strong-coupling
scattering of gluons has an alternate description in terms of strings
moving in a weakly-curved five-dimensional gravitational background.
On the other hand, there is also a {\it weak-weak} duality of some
kind between gravity and gauge theory, directly in four dimensions: It
is possible to write perturbative scattering amplitudes for
gravitons in terms of ``double copies'' of gluon amplitudes.

The original examples of such relations are due
to Kawai, Lewellen and Tye (KLT)~\cite{KLT}, who found them by first deriving
relations between closed and open string theory tree amplitudes, whose
low-energy limits give tree-level relations in field-theory.
Graviton amplitudes are represented in terms of sums of products of
pairs of gluon amplitudes with different color orderings.
The KLT relations hold not only for pure-graviton scattering, in terms
of pure-gauge theory, but also for $\NeqEight$ supergravity (or any 
subsector of it) in terms of amplitudes for $\NeqFour$ sYM.
More recently it has been recognized that other double-copy formulas
exist~\cite{BCJ08,BCJ10}.  The new formulas are simpler in some sense than the
KLT relations; they involve the direct squares of gauge-theory components,
whereas the KLT relations employ multiple permutations of full gauge-theory
amplitudes.  As reviewed in this issue by Carrasco and Johansson~\cite{CJHere},
these relations, combined with generalized unitarity,
have important consequences for multi-loop
amplitudes in $\NeqEight$ supergravity, because they map the problem to
a much simpler one, the corresponding multi-loop amplitudes for
(non-planar) $\NeqFour$ sYM.

In the next two sections we discuss a few of the basic tools used to
organize and illuminate the color and kinematic structure of amplitudes,
which underlay the discussions in many of the other articles in
this special issue.


\section{Color Preliminaries}
\label{ColorPreliminariesSection}

In this section we describe some common conventions for organizing the
color structure of SU$(N_c)$ gauge theory amplitudes, which are used
elsewhere in this special issue.  Most of the articles implicitly discuss
color-ordered partial amplitudes, which are particularly convenient
in the large-$N_c$ limit in which planar diagrams dominate.
Color-ordered amplitudes emerge from a ``trace-based'' color decomposition
(as also reviewed in {\it e.g.}~refs.~\cite{ManganoParke,LDTASI}).
However, in some cases, in particular for describing subleading-color
terms in amplitudes, and color-kinematic duality relations~\cite{CJHere}, 
decompositions based on the SU$(N_c)$ structure constants are more useful.

In general, we consider two different SU$(N_c)$ representations for the
external states:
\begin{itemize}
\item The adjoint representation, for the gluon and any of
its superpartners ({\it i.e.}~for all states in $\NeqFour$ sYM).
Adjoint color indices are denoted by
$a,b,c,a_i,\ldots\in\{1,2,\ldots,N_c^2-1\}$.
\item The fundamental (defining) representation $N_c$, and its conjugate
representation $\overline{N}_c$, for quarks and anti-quarks respectively.
Fundamental color indices are denoted by
$i_1,i_2,\ldots\in\{1,2,\ldots,N_c\}$, and
anti-fundamental $\overline{N}_c$ indices by
$\jb_1,\jb_2,\ldots\in\{1,2,\ldots,N_c\}$.
\end{itemize}
The generators of SU$(N_c)$ in the fundamental representation are traceless
hermitian $N_c\times N_c$ matrices, denoted by $(T^a)_i^{\jb}$.
It is conventional
when discussing helicity amplitudes to normalize the generators by
$\Tr(T^aT^b) = \delta^{ab}$, in order to avoid a proliferation of $\sqrt{2}$'s.

\begin{figure}[t]
\centerline{\epsfxsize 4.8 truein \epsfbox{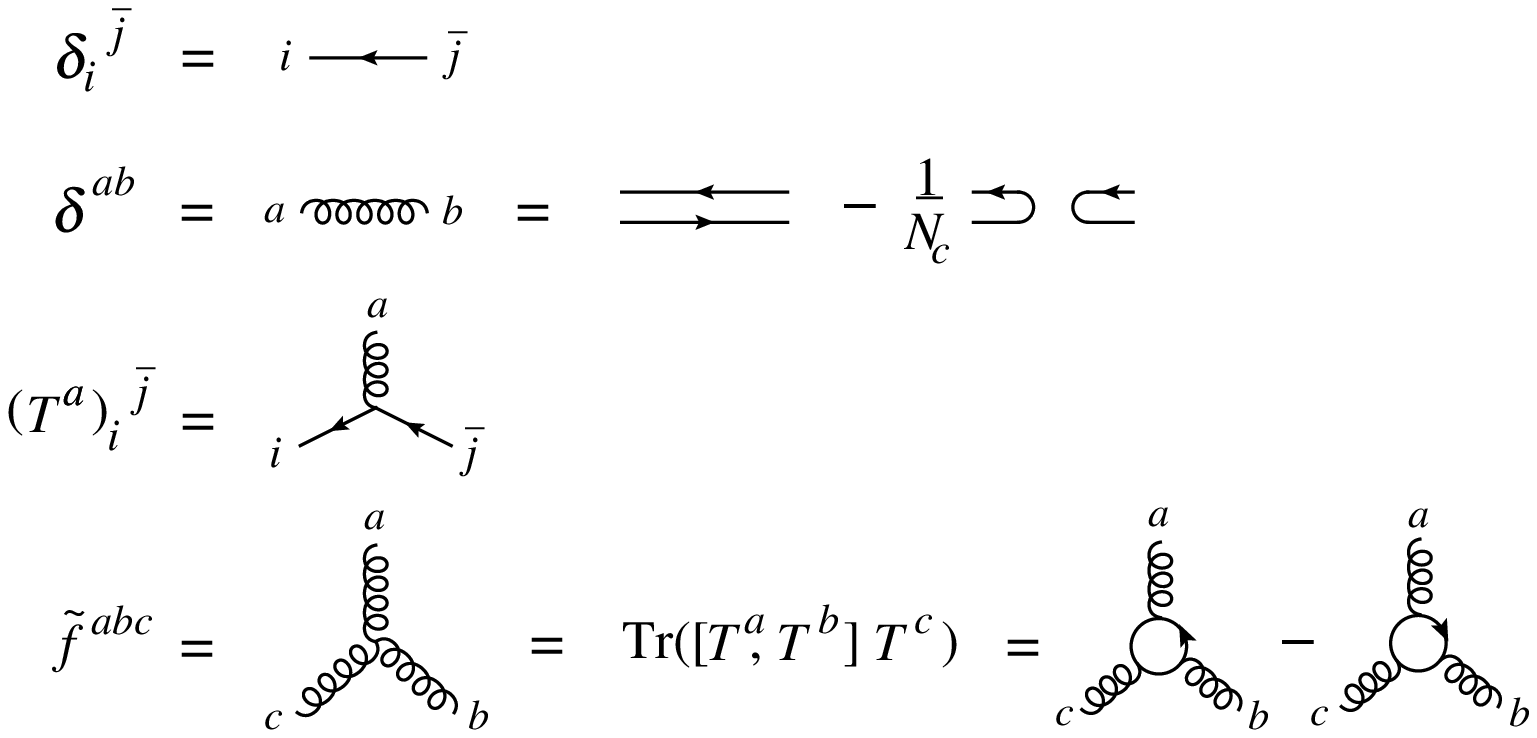}}
\caption[a]{\small Graphical representation of building
blocks for SU$(N_c)$ color factors.}
\label{TfdeltaFigure}
\end{figure}

In QCD, the group theory (color) factors in Feynman diagrams are of
two types: $(T^a)_i^{\jb}$ for the gluon-quark-antiquark vertex,
and the SU$(N_c)$ structure constants $f^{abc}$ (or products of
$f^{abc}$'s) for all other vertices.  (It is also convenient to
normalize the structure constants differently, using
$\tf^{abc} \equiv i\sqrt{2} f^{abc}$, where $f^{abc}$ is the standard,
textbook normalization.)  The color factors 
can be represented diagrammatically~\cite{Cvitanovic}
using Feynman-diagram notation,
as in \fig{TfdeltaFigure}.  Lines with arrows (quarks) carry
fundamental representation indices; curly lines (gluons) carry
adjoint indices.  The adjoint representation is also a bi-fundamental
($N_c\times \overline{N}_c$) representation from which the singlet,
or trace component, has been projected out, as in the second equation
for $\delta^{ab}$ in the figure.  This relation also gives
rise to the SU$(N_c)$ Fierz identity,
\be
  (T^a)_{i_1}^{~\jb_1} \, (T^a)_{i_2}^{~\jb_2}\ =\ 
  \delta_{i_1}^{~\jb_2} \delta_{i_2}^{~\jb_1}
  - {1\over N_c} \, \delta_{i_1}^{~\jb_1} \delta_{i_2}^{~\jb_2}\,,
\label{SUNFierz}
\ee
which allows one to simplify products of traces of $T^a$'s.

\subsection{Trace-based color decompositions}
\label{TraceBasedSubsection}

Color-ordering~\cite{TreeColor} is related to the
't~Hooft double-line formalism~\cite{tHooftColor}.  We begin by
rewriting the color factors entirely in terms of the $T^a$
generators, using the relation
\be
\tf^{abc} \equiv i\sqrt{2}\,f^{abc}= \Tr([T^a,T^b] T^c) \,,
\label{removefabc}
\ee
which follows from the definition of the structure constants, and
is depicted graph\-ically in \fig{TfdeltaFigure}.
After inserting \eqn{removefabc} repeatedly into the color factor
for a typical Feynman diagram, one obtains a large number of 
traces of the generic form 
$\Tr\bigl(\ldots T^a\ldots\bigr)\,\Tr\bigl(\ldots T^a\ldots\bigr)
\,\ldots\,\Tr\bigl(\ldots)$.
If the amplitude has external quark legs, then there will also
be strings of $T^a$'s terminated by fundamental indices,
of the form $(T^{a_1}\ldots T^{a_m})_{i_2}^{~\ib_1}$, one for each
external quark-antiquark pair.

\begin{figure}[t]
\centerline{\epsfxsize 5.0 truein \epsfbox{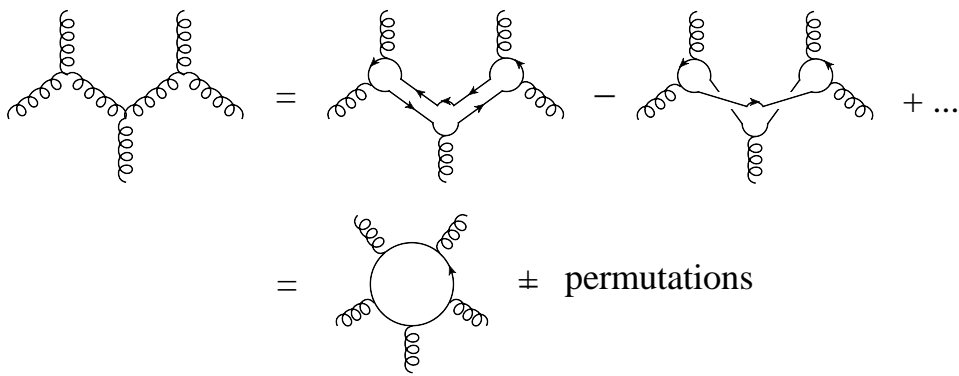}}
\caption[a]{\small Reduction of color factors for $n$-gluon tree amplitudes
to a single trace of $T^a$ generators.}
\label{TreeSingleTraceFigure}
\end{figure}

The number of traces can be reduced considerably by repeated use
of the SU$(N_c)$ Fierz identity~(\ref{SUNFierz}).  In the case
of tree-level amplitudes with external states in the adjoint
representation, such as $n$-gluon amplitudes, \fig{TreeSingleTraceFigure}
sketches how the color factors all may be reduced to a single trace,
$\Tr\bigl(T^{a_{\s(1)}}T^{a_{\s(2)}} \cdots T^{a_{\s(n)}}\bigr)$,
for some permutation $\s\in S_n$ of the $n$ gluons.
This reduction leads to the {\it trace-based color decomposition}
for $n$-gluon tree amplitudes,
\be
\hskip-20mm
 {\cal A}^\tree_n(\{k_i,h_i,a_i\})
 = g^{n-2} \hskip-1.3mm \sum_{\s \in S_n/Z_n} \hskip-1.3mm  
    \Tr(T^{a_{\s(1)}}\cdots T^{a_{\s(n)}})\ 
     A_n^\tree(\s(1^{h_1}),\ldots,\s(n^{h_n}))\,. 
\label{treegluecolor}
\ee
Here ${\cal A}^\tree_n$ is the full amplitude, with dependence
on the external gluon momenta $k_i$, $i=1,2,\ldots,n$,
helicities $h_i = \pm1$, and adjoint indices $a_i$.  In QCD,
the gauge coupling $g$ is related to the strong coupling by
$\alpha_s = g^2/(4\pi)$.  The {\it partial amplitudes}
(or {\it primitive} or {\it color-stripped amplitudes})
$A_n^\tree(1^{h_1},\ldots,n^{h_n})$ have had all the color
factors removed, but contain all the kinematic information.
Cyclic permutations of the arguments of a
partial amplitude, denoted by $Z_n$, leave it invariant, because
the associated trace is invariant under these operations.
However, all $(n-1)!$ non-cyclic permutations, or orderings,
of the partial amplitude appear in \eqn{treegluecolor}.  These
permutations are denoted by $\s\in S_n/Z_n \equiv S_{n-1}$.

Looking again at the way the different trace factors arise in
\fig{TreeSingleTraceFigure}, one sees that the partial amplitude
$A_n^\tree(1,2,\ldots,n)$ only receives contributions from
tree-level Feynman diagrams that can be drawn on a plane, in which
the cyclic ordering of the external legs, $1,2,\ldots,n$,
matches the ordering of the arguments in $A_n^\tree$.
Therefore each partial amplitude can only have singularities
in momentum invariants formed by squaring color-adjacent sums
of momenta, such as $s_{i,i+1} \equiv (k_i + k_{i+1})^2$, 
$s_{i,i+1,i+2} \equiv (k_i + k_{i+1} + k_{i+2})^2$, {\it etc.},
where all indices are defined modulo $n$.
In this way, the color decomposition~(\ref{treegluecolor}) disentangles
the kinematic complexity of the full amplitude ${\cal A}_n^\tree$.

Similarly, tree amplitudes with two external
quarks and $(n-2)$ gluons 
can be reduced to single strings of $T^a$ matrices,
\bea 
 {\cal A}^\tree_n(q_1,g_3,\ldots,g_{n-1},\bar{q}_n) 
 &=& g^{n-2} \hskip-1.3mm \sum_{\s \in S_{n-2}} \hskip-1.3mm  
   (T^{a_{\s(2)}}\cdots T^{a_{\s(n-1)}})_{i_1}^{~\jb_n}
\nonumber\\
&&\hskip1.3cm\null\times
     A_n^\tree(1_q,\s(2),\ldots,\s(n-1),n_{\bar{q}}), 
\label{treequarkgluecolor}
\eea
where we have omitted the helicity labels, and
numbers without subscripts in the argument of $A_n^\tree$
refer to gluons.  In this case there are $(n-2)!$ terms, corresponding
to all possible gluon orderings between the quarks.
Color decompositions for amplitudes containing
additional external quark pairs have also been
described~\cite{ManganoParke}.

We note that the partial amplitudes appearing in \eqn{treequarkgluecolor}
could also be used to describe the scattering of two fermions with
different color quantum numbers, and $(n-2)$ gluons.
For example, if the fermions are gluinos in the adjoint representation,
then the color decomposition has the form of \eqn{treegluecolor},
but the partial amplitudes, with two external fermions, are of the type
given in \eqn{treequarkgluecolor}.  This is easy to see if the two
fermions are color-adjacent; essentially all one needs to do to go
between the two cases is remove one color line running between the two
gluinos.  However, the other cyclic orderings can also
be obtained from $A_n^\tree(1_q,2,\ldots,n-1,n_{\bar{q}})$,
using the Kleiss-Kuijf relations discussed in the next
subsection.  This property illustrates the universality of the
color-ordered primitive amplitudes alluded to in the introduction.

Many articles in this issue consider loop-level amplitudes
in which all of the external states are gluons in the adjoint representation.
The general color decomposition here is similar to \eqn{treegluecolor},
except that multiple color traces are now possible.  Roughly speaking,
a loop of gluons carries a fundamental and an anti-fundamental index
around the loop with it, and an external gluon can attach to either index.
The number of traces at $L$ loops is equal to $L+1$.

At one loop, the full color decomposition for $n$ gluons
is~\cite{BKLoopColor}
\bea
\hskip-24mm
{\cal A}^\oneloop_n(\{k_i,h_i,a_i\})
&=& g^n\Biggl[
    \sum_{\s \in S_n/Z_n} 
    N_c\,\Tr(T^{a_{\s(1)}}\cdots T^{a_{\s(n)}})\ 
     A_{n;1}(\s(1^{h_1}),\ldots,\s(n^{h_n})) 
\nonumber \\
&& \hskip 5mm 
 +\ \sum_{c=2}^{\lfloor{n/2}\rfloor+1}
      \sum_{\s \in S_n/S_{n;c}}
    \Tr(T^{a_{\s(1)}}\cdots T^{a_{\s(c-1)}})\ 
    \Tr(T^{a_{\s(c)}}\cdots T^{a_{\s(n)}})\ 
\nonumber \\
&& \hskip 43mm
 \times\ A_{n;c}(\s(1^{h_1}),\ldots,\s(n^{h_n}))
     \Biggr]\,. 
\label{loopgluecolor}     
\eea
Here $A_{n;c}$ are the partial amplitudes, 
$Z_n$ and $S_{n;c}$ are the subsets of $S_n$
that leave the corresponding single and double trace structures 
invariant, and $\lfloor x \rfloor$ is the greatest integer less than or
equal to $x$.  The formula is more complicated than at tree level, but
only by the need to keep track of more distinct trace structures and
their various symmetries.  

If there are $n_f$ flavors of quarks in the loop,
it is easy to see that they only contribute to the single-trace
coefficient $A_{n;1}$ in \eqn{loopgluecolor}, and with a
weight $n_f/N_c$ relative to the adjoint gluons.
(Similar color decompositions are available for one-loop amplitudes
with an external quark pair~\cite{BDK2f3g}.)

When one constructs the color-summed cross section,
as is usually required for QCD applications,
the contribution of the double-trace coefficients,
$A_{n;c}$ for $c>1$, is suppressed by a power of $1/N_c^2$ with respect
to that of $A_{n;1}$.  (It also turns out that the $A_{n;c}$ are
not really independent, but can be computed as a sum over permutations
of the $A_{n;1}$~\cite{BDDKNeq4}.)  Thus \eqn{loopgluecolor}
simplifies a lot in the large-$N_c$ limit.
The factor of $N_c = \Tr(1)$ simply comes from those terms in which all 
external gluons attach to the same fundamental index line, leaving the
trace of the identity matrix for the untouched line.

Correspondingly, if we only want the leading terms in the large-$N_c$
limit at $L$ loops, we can write the compact color decomposition,
\bea
&&\hskip-1.5cm
{\cal A}^\Lloop_n(\{k_i,h_i,a_i\})|_{\rm leading\hbox{-}color}
\nonumber\\
&&\hskip-0.5cm
= g^{n-2} \, (g^2 N_c)^L 
    \sum_{\s \in S_n/Z_n} 
    \Tr(T^{a_{\s(1)}}\cdots T^{a_{\s(n)}})\ 
     A_{n}^{(L)}(\s(1^{h_1}),\ldots,\s(n^{h_n}))\,,
\label{Lloopleadingcolor}     
\eea
where we have dropped the ``;1'' index on the leading-color partial
amplitude.  The 't~Hooft coupling $\lambda = g^2 N_c$ emerges naturally
in this limit.  The articles in this issue about multi-loop amplitudes
in planar $\NeqFour$~sYM are generally concerned with
the $A_n^{(L)}$ in \eqn{Lloopleadingcolor}.  
Because each loop integration typically brings a factor of $1/(4\pi)^2$
(or $1/(4\pi)^{2-\e}$ when using dimensional regularization in $D=4-2\e$),
in some cases the precise normalizations 
of the $A_n^{(L)}$ will differ by this factor,
as well as possibly other conventional factors.

\subsection{$f$-based color decompositions}
\label{FBasedSubsection}

\begin{figure}[t]
\centerline{\epsfxsize 4.0 truein \epsfbox{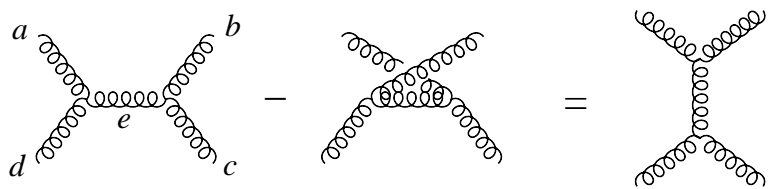}}
\caption[a]{\small The Jacobi identity~(\ref{ColorJacobiIdentity}).}
\label{ColorJacobiFigure}
\end{figure}

There is another type of color decomposition that is actually more useful
than the above trace-based decomposition for addressing certain issues,
and particularly for working beyond the large-$N_c$ approximation.
We could call this approach the $f$-based decomposition, because it goes
back to the original color factors built out of the structure constants
$f^{abc}$.  (Now we assume that all states are in the adjoint representation.)
The diagrammatic representation of a general $f$-based color factor at
tree level has the structure of a cubic tree graph, with
vertices given by $\tf^{abc}$'s and propagators given by $\delta^{ab}$'s.
However, these color factors are not all independent, due to the color
Jacobi identity.  This identity can be written in different ways.
One way,
\be
\tf^{dae} \tf^{bce} - \tf^{dbe} \tf^{ace} = \tf^{abe} \tf^{cde} \,,
\label{ColorJacobiIdentity}
\ee
which is also depicted in \fig{ColorJacobiFigure}, corresponds to
the fact that the structure constants are also the SU$(N_c)$ generators
in the adjoint representation, $\tf^{abc} = (F^b)_{ac}$,
so that their commutator gives $\tf^{abe}$ contracted with $F^e$.

At tree level, one can use the identity~(\ref{ColorJacobiIdentity})
repeatedly to transform all $f$-based color factors into a
``multi-peripheral'' form in which two selected gluons, say 1 and $n$,
are always at the end of a long chain of structure constants,
and the other $(n-2)$ gluons
are emitted from along the ladder~\cite{DFMColor,DDMColor},
\bea
\hskip-20mm
 {\cal A}_n^\tree (\{a_i\}) &=&    
  g^{n-2} \sum_{\s\in S_{n-2}}
      \tf^{a_1 a_{\s(2)} x_1} \tf^{x_1 a_{\s(3)} x_2} \cdots
   \tf^{x_{n-3} a_{\s(n-1)} a_n}\nonumber\\
&&\null\hskip20mm \times
    A_n^\tree(1,\s(2),\ldots,\s(n-1),n)\, \nonumber\\
&=& g^{n-2} \sum_{\s\in S_{n-2}}
      ( F^{a_{\s(2)}} \cdots F^{a_{\s(n-1)}} )_{a_1 a_n}
      A_n^\tree(1,\s(2),\ldots,\s(n-1),n)\,.
\label{GluonDecompNew}
\eea
Notice that this representation is identical in form to the decomposition
for two quarks and $(n-2)$ gluons in \eqn{treequarkgluecolor}, except
for the representation used for the SU$(N_c)$ generator matrices.
Although \eqns{treequarkgluecolor}{GluonDecompNew} correspond to the
cases of fundamental and adjoint representations, respectively,
the same type of decomposition clearly holds for two external matter fields
in an arbitrary SU$(N_c)$ representation.  The fact that the coefficients
of the $f$-based decomposition~(\ref{GluonDecompNew}) are identical 
to the color-ordered partial amplitudes $A_n^\tree$ appearing in the
trace-based decomposition can be demonstrated by
contracting both representations with $\Tr(T^{a_1}T^{a_2}\cdots T^{a_n})$ and
keeping only the leading terms at large $N_c$~\cite{DDMColor}.

Note that there are only $(n-2)!$ terms in the $f$-based color
decomposition, in contrast to the $(n-1)!$ terms in the trace-based
decomposition~(\ref{treegluecolor}).  Therefore there are identities
between the $A_n^\tree$, which are group-theoretical in nature ---
generalizations of certain $U(1)$ decoupling
identities~\cite{ManganoParke,LDTASI}.
These identities can be used to put any two legs next
to each other, for example legs 1 and $n$:
\be
A_n^\tree(1,\{\alpha\},n,\{\beta\}) =
 (-1)^{n_\beta} \sum_{\s \in {\rm OP}\{\alpha\}\{\beta^T\}}
 A_n^\tree(1,\s(\{\alpha\}\{\beta^T\}),n) \,,
\label{KKequation}
\ee
where $\{\alpha\}$ and $\{\beta\}$ are two sets of external gluons,
whose union is $\{2,3,\ldots,n-1\}$.  Also, $n_\beta$ is the
number of elements in the set $\{\beta\}$, the set $\{\beta^T\}$ is
$\{\beta\}$ with the ordering reversed, and ${\rm OP}\{\alpha\}\{\beta^T\}$ is
the set of ordered permutations, or ``mergings'', of the 
two sets $\{\alpha\}$ and $\{\beta^T\}$ that preserve the 
ordering of the $\alpha_i$ within $\{\alpha\}$ and of the $\beta_i$ 
within $\{\beta^T\}$, while allowing for all possible relative orderings 
of the $\alpha_i$ with respect to the $\beta_i$.  These relations
are known as the Kleiss-Kuijf relations~\cite{KK}.  They can be derived
from \eqn{GluonDecompNew} by expanding out the $\tf^{abc}$
factors using \eqn{removefabc}, identifying the coefficients of the
trace structures, and comparing them with the original trace-based
decomposition~(\ref{treegluecolor})~\cite{DDMColor}.

In fact, it has been realized recently that there are not even $(n-2)!$
independent color-ordered amplitudes $A_n^\tree$, but only
$(n-3)!$~\cite{BCJ08}, as reviewed in this issue by Carrasco and
Johansson~\cite{CJHere}. The additional linear relations between $A_n^\tree$'s
are not purely group-theoretical in nature, but also involve kinematical
factors.  They allow one to put any {\it three} external gluon legs next to
each other. For example, for legs 1, 2 and $n$, the identities take
the form,
\be
\hskip-1.0cm
A_n^\tree(\s(1),\s(2),\ldots,\s(n))
= \sum_{\rho\in S_{n-3}} K^{(\s)}_\rho \, 
A_n^\tree(1,2,\rho(3),\ldots,\rho(n-1),n) \,,
\label{BCJ3}
\ee
where $\s$ and $\rho$ are permutations and $K^{(\s)}_\rho$ are
kinematic-dependent (but not state-dependent)
coefficients~\cite{BCJ08,CJHere}.
These relations were first proved using open string theory~\cite{BCJStringProof},
using a contour deformation argument~\cite{Plahte} similar to that used
in deriving the KLT relations~\cite{KLT} between gravity and gauge theory
amplitudes.  They have also been proven directly in field theory, using
a recursive argument~\cite{BCJRecursiveProof}.

Although \eqn{BCJ3} expresses the consequences of certain color-kinematic
identities in the trace basis, the identities themselves are more
naturally stated in an $f$-based approach.  For this purpose it is useful to
{\it not} transform the $f$-based color factors into multi-peripheral form, 
but leave the color decomposition free, in terms of the set of all cubic
graphs $\Gamma_3$, writing
\be
{\cal A}_n^\tree = g^{n-2} \sum_{i\in\Gamma_3} 
\frac{N_i C_i}{ (\prod_j p_j^2)_i } \,.
\label{BCJtreerep}
\ee
The color factors $C_i$ are products of structure constants for each vertex
$V_i$ of the $i^{\rm th}$ graph, contracted together along the propagators,
$C_i = \prod_{j\in V_i} \tf^{a_{j_1}a_{j_2}a_{j_3}}$.
The factors of $1/p_j^2$ are scalar propagators, one for each internal line
in the graph, while the $N_i$ are kinematical numerator factors.
Consider a triplet of graphs in $\Gamma_3$, which are identical except for
a region from which four lines emanate.  Within this region they differ
according to the three ways of joining four legs with two cubic vertices,
shown in \fig{ColorJacobiFigure}.  Call the three graphs $s$, $t$ and
$u$, and let the color factors $C_{s,t,u}$ be normalized (signed) so that they
obey the Jacobi identity~(\ref{ColorJacobiIdentity}) as $C_s - C_t = C_u$.
Then the statement of color-kinematic duality~\cite{BCJ08} is that the
associated numerator factors should be related by $N_s - N_t = N_u$,
for every such triplet of graphs.  A simple and intuitive argument for
these relations has been provided based on the heterotic
string~\cite{TyeZhang}.

The existence of such a representation with {\it local} $N_i$
({\it i.e.}~polynomials in the momenta) has been checked in various examples,
not only at tree level, but also through three loops for $\NeqFour$
sYM amplitudes~\cite{BCJ10}.  As reviewed here by Carrasco and
Johansson~\cite{CJHere},
the existence of a representation for gauge theory amplitudes satisfying
$N_s - N_t = N_u$ for each triplet of Jacobi-related cubic graphs
has important implications.  Large classes of numerator factors are
related to each other, and gravitational amplitudes are easily constructed
from gauge theory ones.


\section{Kinematic Preliminaries}
\label{KinematicPreliminariesSection}

Now that we have described two different ways to organize the color quantum
numbers for scattering amplitudes, let us turn to
kinematical issues, including
convenient choices of external states and kinematical variables.

\subsection{Spinor-helicity formalism}
\label{SpinorHelicitySubsection}

In QCD, the helicities of massless quarks are conserved by their
chirality-preserving interactions with gluons.  Hence it is natural to
use a helicity basis for the quark amplitudes.  In four-component
notation, the spinors for the external states are taken to be
$u_\pm(k) = \frac{1}{2}(1\pm\gamma_5)u(k)$
for a quark with momentum $k$,
and $v_\mp(k)=\frac{1}{2}(1\mp\gamma_5)v(k)$ for an anti-quark.
However, in the massless limit these spinors can be chosen to be equal
to each other, $u_\pm(k)=v_\mp(k)$, and we can use two-component Weyl
spinors as well.  We will see that helicity amplitudes for external
gluons can be built from the same objects.

We want to consider amplitudes with $n$ different
momenta $k_i$, $i=1,2,\ldots,n$, and to do so in a way that respects
crossing symmetry.  Thus we take all the momenta to be outgoing,
so that momentum conservation reads $\sum_i k_i^\mu = 0$.  In a realistic
process with some incoming particles, the physical momentum for each of the 
incoming particles is simply the {\it negative} of the corresponding $k_i$.
Also, the physical helicity $h_i$ is the negative of the one by which 
we label the amplitude. (The spin $S_i$ does not reverse under crossing,
but $h_i = S_i\cdot k_i$ does.)  The two-component spinor for the
$i^{\rm th}$ state, if it is an outgoing fermion with helicity
$\pm\frac{1}{2}$, has several notations~\cite{ManganoParke,LDTASI,ItaHere}:
\bea
(\lambda_i)_\alpha\ &\equiv&\ |i^+\rangle\ \equiv\ |k_i^+\rangle\
\equiv\ [u_+(k_i)]_\alpha\ =\ [v_-(k_i)]_\alpha \,, \\
(\tilde\lambda_i)_{\dot\alpha}\ &\equiv&\ |i^-\rangle\ \equiv\ |k_i^-\rangle
\equiv\ [u_-(k_i)]_{\dot\alpha}\ =\ [v_+(k_i)]_{\dot\alpha} \,.
\eea
Similarly, the conjugate spinors are
\bea
(\tilde\lambda_i)^{\dot\alpha}\ &\equiv&\ \langle i^+|\ \equiv\ \langle k_i^+|\
\equiv\ [\overline{u_+(k_i)}]^{\dot\alpha}\
=\ [\overline{v_-(k_i)}]^{\dot\alpha} \,, \\
(\lambda_i)^\alpha\ &\equiv&\ \langle i^-|\ \equiv\ \langle k_i^-|\
\equiv\ [\overline{u_-(k_i)}]^\alpha\ =\ [\overline{v_+(k_i)}]^\alpha \,.
\eea
Two-component indices are raised and lowered with the two-dimensional
antisymmetric tensors $\eps^{\alpha\beta}$, $\eps^{\dot\alpha\dot\beta}$,
{\it etc.}  The spinor inner products come from contracting spinors
for different momenta with these tensors,
\bea
\spa{j}.{l}\ &=&\ 
\eps^{\alpha\beta} (\lambda_j)_\alpha (\lambda_l)_\beta
\ =\ \langle j^- | l^+\rangle\ =\ \overline{u_-(k_j)} u_+(k_l) \,,
\label{spadef} \\
\spb{j}.{l}\ &=&\ 
\eps^{\dot\alpha\dot\beta} (\tilde\lambda_j)_{\dot\alpha}
                           (\tilde\lambda_l)_{\dot\beta}
\ =\ \langle j^+ | l^-\rangle\ =\ \overline{u_+(k_j)} u_-(k_l) \,.
\label{spbdef}
\eea
Spinor products are antisymmetric under exchange of labels,
$\spa{j}.{l}=-\spa{l}.{j}$, $\spb{j}.{l}=-\spb{l}.{j}$,
and $\spa{j}.{j}=\spb{j}.{j}=0$.  Lorentz vectors can be written as 
bi-spinors, or $2\times 2$ matrices, by contracting them
with the Pauli matrices.  A massless vector written in this way factorizes
into the product of two massless spinors, as
\be
(\ksl_i)_{\alpha\dot\alpha}\ \equiv\
k_i^\mu (\sigma_\mu)_{\alpha\dot\alpha}
\ =\ u_+(k_i) \overline{u_+(k_i)}
\ =\ (\lambda_i)_\alpha (\tilde\lambda_i)_{\dot\alpha} \,.
\label{masslesskfact}
\ee
This factorization plays a role in the BCFW complex-momentum
shift, which is best described in terms of the $\lambda$ and
$\tilde\lambda$ variables~\cite{BCFW}, as is done elsewhere in this
issue~\cite{BSTHere,ItaHere,DrummondHere}.

Amplitudes with external gluons can also
be described in terms of the $\lambda_i$ and $\tilde\lambda_i$ variables,
thanks to the spinor-helicity formalism~\cite{SpinorHelicity}.
The polarization vector $\pol^\pm(k)$ for an outgoing massless vector
particle with momentum $k$ and helicity $h=\pm1$, is required to be
transverse to $k$, $\pol^\pm\cdot k = 0$.  In the spinor-helicity formalism,
$\pol^\pm$ is also chosen to be transverse to another
massless momentum $q$, called the {\it reference momentum} (which should
not be parallel to $k$ but is otherwise arbitrary).
Note that $k$ and $q$ span a two-dimensional subspace of four-dimensional
space-time.  Because $\ksl |k^\pm\rangle = \qsl | q^\pm\rangle = 0$,
the two orthogonal directions to this subspace are spanned by
$\langle q^+ | \gamma^\mu | k^+\rangle$ and
$\langle q^- | \gamma^\mu | k^-\rangle$.  The spinor-helicity polarization
vectors live in this subspace; indeed, they are proportional to these
two vectors, and are normalized by the condition that 
$\pol^\pm \cdot (\pol^\pm)^* = -1$:
\be
\pol_\mu^\pm(k,q) = \pm\frac{\langle q^\mp | \gamma_\mu | k^\mp\rangle}
{\sqrt{2} \langle q^\mp | k^\pm \rangle} \,.
\label{spinorheldef}
\ee
As a bi-spinor, $\pol^\pm(k_i,q_i)$ is given by
\be
\hskip-1cm
[\pol^+(k_i,q_i)]_{\alpha\dot\alpha}
= \sqrt{2} 
\frac{(\lambda_{q_i})_\alpha (\tilde\lambda_{i})_{\dot\alpha}}
     { \spa{q_i}.{i} } \,,
\qquad
[\pol^-(k_i,q_i)]_{\alpha\dot\alpha}
= - \sqrt{2}
\frac{(\lambda_{i})_\alpha (\tilde\lambda_{q_i})_{\dot\alpha}}
     { \spb{q_i}.{i} } \,.
\label{spinorheldef2}
\ee
If each $q_i$ is chosen to be another momentum in the process, say $k_j$,
then it is clear from the general form of the 
Feynman rules (using also \eqn{masslesskfact})
that the full amplitude can be built entirely out
of the spinor products $\spa{j}.{l}$ and $\spb{j}.{l}$, for
$1\leq j,l \leq n$.  

What are the basic properties of the spinor products which make them
so convenient variables for helicity amplitudes?
First of all, for real momenta, the two types of spinor products
are complex conjugates of each other, $\spa{j}.{l}=\pm{\spb{l}.{j}}^*$.
Also, for either real or complex momenta they ``square'' to give the
ordinary dot products of momenta,
\be
\spa{l}.{j}\spb{j}.{l}
\ =\ \Tr[\textstyle{\frac{1}{2}}(1-\gamma_5) \ksl_l \ksl_j]
\ =\ 2 k_j\cdot k_l = s_{jl} \,.
\label{sfromspinors}
\ee
(\Eqn{sfromspinors} is written in the convention usually used in the
QCD literature, but the reader should be aware that another convention
exists, in which the sign of $\spb{j}.{l}$ is reversed so
that $\spa{l}.{j}\spb{j}.{l} = -s_{jl}$.)  

\begin{figure}[t]
\centerline{\epsfxsize 4.5 truein \epsfbox{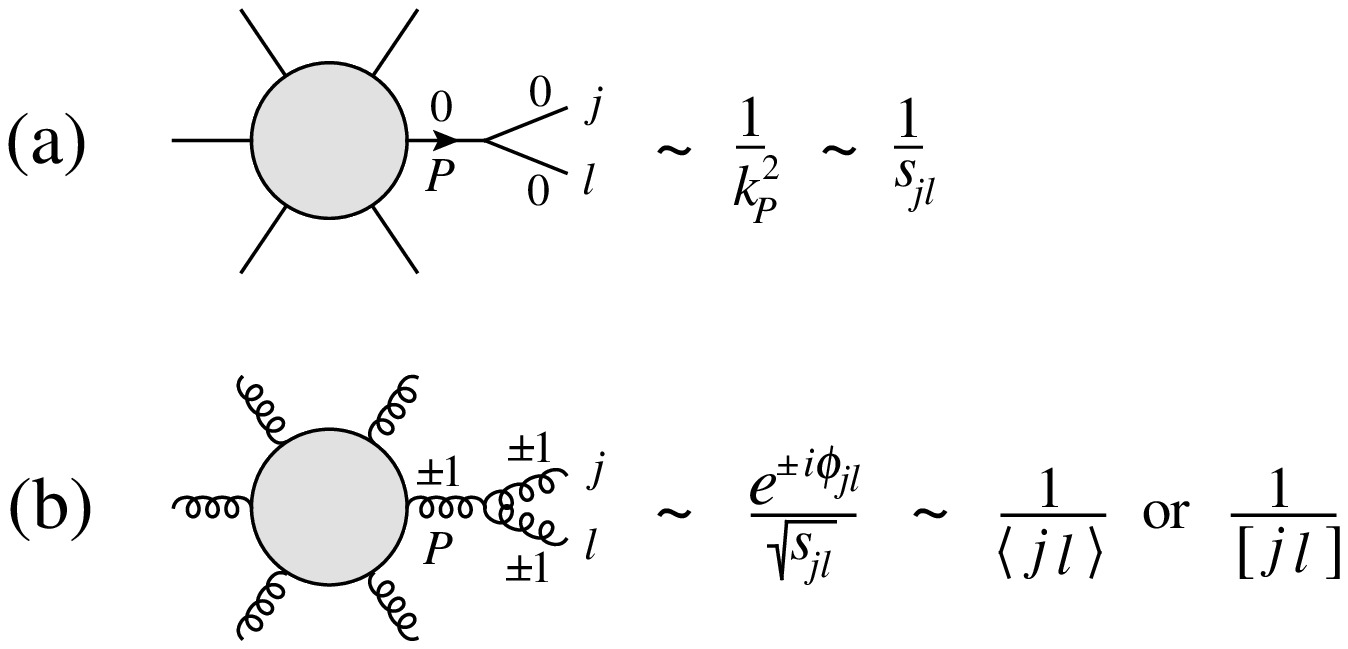}}
\caption[a]{\small (a) A typical collinear limit in a theory with
massless scalars.  There is no suppression of the singularity
from angular-momentum conservation.  (b) A typical collinear limit
in a massless gauge theory, for which there is always a mismatch of
at least one unit of angular momentum, lessening the singularity.}
\label{CollLimitFigure}
\end{figure}

For real momenta, the conjugation property and \eqn{sfromspinors}
imply that
\be
\spa{j}.{l}\ =\ \sqrt{s_{jl}} \, e^{i\phi_{jl}} \,, \qquad
\spb{j}.{l}\ =\ \sqrt{s_{jl}} \, e^{-i\phi_{jl}} \,,
\label{realmomsqrt}
\ee
where $\phi_{jl}$ is a phase.  This phase shifts as $k_j$ and $k_l$
are rotated in azimuthal angle
around the axis corresponding to their sum, $k_P = k_j + k_l$.
One way of seeing the utility of the spinor products for helicity amplitudes
is to examine the limits in which two particles become collinear,
$k_P^2 = s_{jl} \to 0$.  \Fig{CollLimitFigure}(a) shows that the typical
behavior of amplitudes in a massless scalar field theory is
$\sim 1/k_P^2 = 1/s_{jl}$, coming just from the scalar propagator.
However, in massless gauge theory, there are numerator
factors in the Feynman diagrams which lessen this divergence.
More physically, the factorization in the collinear limit should be 
onto a physical gluon state with helicity $\pm1$.  However, the sum
of the two external helicities must be $\pm2$ or 0.
Therefore there is at least a $\pm1$ mismatch of the spin angular momentum
along the $k_P$ axis, as illustrated in \fig{CollLimitFigure}(b).
The spin angular-momentum mismatch requires some orbital angular momentum,
which in turn causes a suppression of the magnitude of the
amplitude, from $1/s_{jl}$ to $1/\sqrt{s_{jl}}$.  It also dictates
a phase shift under azimuthal rotation of $k_j$ and $k_l$ about the
$k_P$ axis.  Both of these properties are captured by the spinor products,
making them ideal for describing helicity amplitudes.

The simplest nonvanishing tree-level gluon amplitudes are
the Parke-Taylor or MHV amplitudes~\cite{ParkeTaylor},
\be
A_n^\tree(1^+,\ldots,l^-,\ldots,m^-,\ldots,n^+) =
i \, { {\spa{l}.{m}}^4 \over \spa1.2\spa2.3\cdots\spa{n}.1 } \,.
\label{MHVPT}
\ee
Here exactly two gluons, $l$ and $m$, have negative helicity, and the
remaining $(n-2)$ gluons have positive helicity.  The amplitudes with
zero or one negative gluon helicity vanish by supersymmetry
Ward identities~\cite{SWI,SWIQCD}.  The only denominator factors in
these amplitudes are the spinor products for color-adjacent pairs
of momenta, $\langle i,\,i+1\rangle$.  They capture such universal
collinear limits~\cite{ManganoParke,BDDKNeq4} as,
\bea
&&\hskip-1cm
A_n^\tree(1^+,\ldots,l^-,\ldots,m^-,\ldots,(n-1)^+,n^+)|_{n-1\parallel n}
\label{collimitexample} \\
&\sim& \frac{1}{\sqrt{z(1-z)} \langle n-1,\,n\rangle} \,
A_{n-1}^\tree(1^+,\ldots,l^-,\ldots,m^-,\ldots,P^+) \,,
\eea
where legs $(n-1)$ and $n$ are becoming parallel, with  
$k_P = k_{n-1} + k_n$, $k_{n-1} \approx z k_P$ and
$k_{n} \approx (1-z) k_P$.


\subsection{Three-point amplitudes and complex kinematics}
\label{ThreePointSubsection}

Another way to see why the spinor products are so useful is to consider
the special case $n=3$ for the MHV amplitudes,
\be
A_3^\tree(1^-,2^-,3^+) = i \, { {\spa{1}.{2}}^4 \over \spa1.2\spa2.3\spa3.1 }
\,. \label{mmp}
\ee
One might think that the scattering of three massless particles is singular,
because all of the $s_{jl}$ vanish since they are equal to the squared
momentum of the remaining particle.  Indeed, for real momenta,
\eqn{realmomsqrt} implies that all the spinor products vanish too, and there
is no way to build a nonvanishing and nonsingular amplitude.
However, for complex momenta, the complex-conjugation relation
$\spa{j}.{l}=\pm{\spb{l}.{j}}^*$ no longer holds, and it is possible for
half the spinor products to be nonvanishing, while the other half vanish.

Specifically, we can choose the three negative-helicity two-component spinors
to be proportional~\cite{WittenTwistor},
\be
\tilde\lambda_1^{\dot\alpha}\ \propto\ \tilde\lambda_2^{\dot\alpha}
\ \propto\ \tilde\lambda_3 ^{\dot\alpha} \,.
\label{spbvanish}
\ee
Then according to \eqn{spbdef} we have $\spb1.2=\spb2.3=\spb1.3=0$.
However, the other three spinor products, $\spa1.2$, $\spa2.3$ and $\spa1.3$,
are allowed to be nonzero. (This is consistent with the vanishing
of \eqn{sfromspinors}, which still holds for complex momenta.)
For this choice of complex kinematics,
the MHV three-point amplitude~(\ref{mmp}) is nonvanishing and nonsingular,
while the conjugate ``$\overline{{\rm MHV}}$'' amplitude,
\be
A_3^\tree(1^+,2^+,3^-) = -i \, { {\spb1.2}^4\over\spb1.2\spb2.3\spb3.1 } \,,
\label{ppm}
\ee
vanishes.  In contrast, for the conjugate kinematics with
\be
\lambda_1^{\alpha}\ \propto\ \lambda_2^{\alpha}
\ \propto\ \lambda_3 ^{\alpha} \,,
\label{spavanish}
\ee
\eqn{mmp} vanishes while \eqn{ppm} is nonvanishing and nonsingular.

The three-point amplitudes~(\ref{mmp}) and (\ref{ppm}), as well as related
amplitudes with two matter particles and one gluon, are very important
in massless gauge theory.  The BCFW relations build all tree-level
scattering amplitudes recursively from them, using factorization
onto multi-particle poles, as well as onto complex-momentum versions
of the collinear poles shown in \fig{CollLimitFigure}.
Of course Feynman diagrams can also be used to build amplitudes from
three- and four-point vertices.  However, Feynman vertices are typically
evaluated with off-shell lines emanating from them,
which makes them gauge-dependent.  In contrast, \eqns{mmp}{ppm} are
on shell, albeit for complex momenta, and therefore they are
fully gauge invariant.  In the recursive construction of $n$-gluon
amplitudes, the four-gluon vertex is never needed, essentially because
it is related to the three-vertex by gauge transformations.

Another reason why the three-point amplitudes are important is for
building up loop amplitudes through generalized unitarity, as reviewed in
this issue by Ita~\cite{ItaHere} and Britto~\cite{BrittoHere} at one loop, and
by Bern and Huang~\cite{BernHuangHere} and Carrasco and Johansson~\cite{CJHere}
at the multi-loop level.  Typically, generalized unitarity cuts place
many propagators on shell, so they very often contain at least one
three-point tree amplitude.  Finally, the relations between gravity and
gauge theory tree amplitudes are the simplest of all for the
three-point case, where they can be written as an exact square,
\be
M_3^\tree(1,2,3) = [ A_3^\tree(1,2,3) ]^2 \,,
\label{KLT3}
\ee
ignoring overall coupling-constant factors.


\subsection{Kinematic variables for planar theories}
\label{PlanarVariablesSubsection}

For planar (large $N_c$) gauge theory amplitudes --- which includes
all tree-level amplitudes --- new sets of kinematic variables have
proved very useful for identifying new structures and symmetries.

\begin{figure}[t]
\centerline{\epsfxsize 4.5 truein \epsfbox{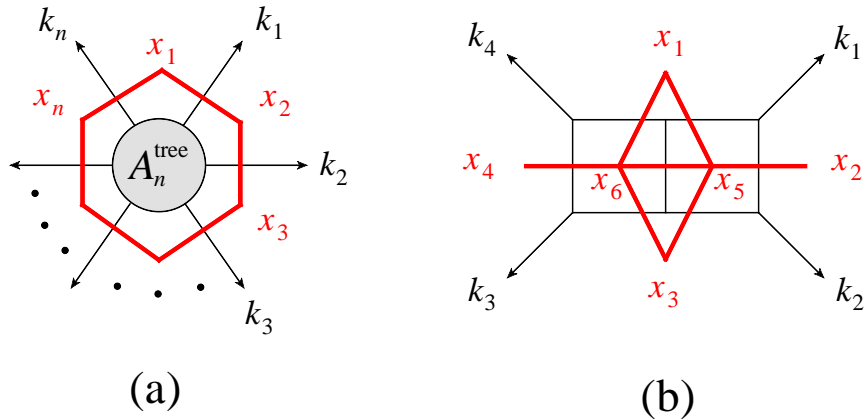}}
\caption[a]{\small (a) Dual coordinates $x_i$ for a color-ordered tree
amplitude with external momenta $k_i$.  The dual variables live in 
the sectors, or regions, demarcated by the $k_i$.
Thin (black) lines denote ordinary momenta, while thick (red) lines
denote differences $x_{i,j}$ between dual coordinates.
These differences also equal the momenta carried by the lines they cross.
(b) Dual coordinates for a particular integral entering
the planar two-loop four-gluon amplitude.  Now additional dual coordinates,
$x_5$ and $x_6$, live within the loops.  In this case we only show the seven
dual coordinate separations associated with the seven propagators for
this integral ($x_{1,5}$, $x_{2,5}$, $x_{3,5}$, $x_{1,6}$, $x_{3,6}$,
$x_{4,6}$ and $x_{5,6}$).}
\label{DualCoordinateFigure}
\end{figure}

Because the external legs have a definite cyclic order in the planar case,
they carve the plane into sectors.  To each such sector, one can assign a 
vertex $x_i^\mu$ for a dual graph, as shown in \fig{DualCoordinateFigure}(a).
Lines connecting vertices of the dual graph cross lines of the original graph.
The latter lines carry momenta.  Differences between $x_i$'s are computed
according to the net momentum flowing through the lines of the original
graph.  Thus in \fig{DualCoordinateFigure}(a), adjacent sectors are separated
by $x_{j+1} - x_{j} = k_j$, or more generally, 
\be
x_{i,j}^\mu \equiv x_{i}^\mu - x_{j}^\mu
= k_j^\mu + k_{j+1}^\mu + \cdots + k_{i-1}^\mu \,.
\label{sectorvariable}
\ee
The $x_i$ are often called {\it dual coordinates} (or sometimes region or
sector variables), but they are essentially momenta, not coordinates.
In a planar loop graph we can assign additional $x_i$ variables to the
faces inside each loop, corresponding to the independent loop momenta.
\Fig{DualCoordinateFigure}(b) shows a two-loop example.
Here the thick lines of the dual graph that are shown
each cross (and therefore correspond to) one propagator of the planar
double box integral.

More precisely, differences of the $x_i$ are momenta.  The $x_i$
themselves are not constrained by momentum conservation.  They are
constrained by on-shell conditions, $x_{j+1,j}^2=k_j^2=0$.
However, unlike momenta, it is possible to invert the $x_i$'s,
according to
\be
 x_i^\mu \to \frac{x_i^\mu}{x_i^2} \,, \qquad
x_{i,j}^2 \to \frac{x_{i,j}^2}{x_i^2 x_j^2} \,. 
\label{dualinversion}
\ee
Remarkably, this inversion is a symmetry of both integrands and
amplitudes in planar $\NeqFour$ sYM.  The integrands are
functions of the combinations $x_{i,j}^2$, which are manifestly
invariant under Lorentz transformations and translations of the $x_i$'s.
(Translation invariance is guaranteed simply because the amplitude depends
on momenta, which are differences of $x_i$'s.)  Therefore
the inversion~(\ref{dualinversion}) combines with
Poincar\'e invariance to generate the conformal group SO(4,2).
This group acts on the dual variables $x_i$, not on space-time
coordinates; hence the symmetry is referred to as dual conformal invariance.

At the loop level, dual conformal invariance can be spoiled by
the infrared regulator needed for on-shell amplitudes.  A loop-integration
measure in four dimensions, $d^4x_i$, transforms under an
inversion~(\ref{dualinversion}) as, $d^4x_i \to d^4x_i/(x_i^2)^4$.
In $D=4$ this factor precisely cancels terms from inverting the associated
propagators in graphs such as the one in \fig{DualCoordinateFigure}(b).
However, in dimensional regularization, we have $D=4-2\eps$, and the
invariance is lost due to extra factors in the transformation
of the loop integration measure.  As discussed in detail
in this issue by Henn~\cite{HennHere}, for planar $\NeqFour$ sYM it
is possible to retain dual conformal invariance at the loop level
by using instead a Higgs regulator~\cite{AHPS},
which generates particle masses that can also be thought of as extra
components of the $x_i^\mu$.  Although amplitude computations with the
Higgs regulator have not been pushed quite as far yet as in
dimensional regularization, there are certain advantages to
this approach.  For example, in dimensional regularization, 
terms in lower-loop amplitudes that vanish as $\eps\to0$ often have to be
computed, because they can multiply pole terms, $\sim 1/\eps^k$, in other
amplitudes (or in the squares of amplitudes needed to construct
differential cross sections).  The article by Schabinger in this issue
discusses such ${\cal O}(\e)$ contributions~\cite{SchabingerHere}. 
In contrast, with the Higgs regulator, the singularities
have the form of powers of logarithms, $\sim \ln^k m^2$, and power-suppressed
terms of the form $m^2/s_{i,i+1}$ can always be dropped, because
they vanish faster than any power of a logarithm.

How constraining is dual conformal symmetry?  In general it can only determine
amplitudes up to arbitrary functions of the cross-ratios,
\be
u_{ijkl}\ \equiv\ \frac{x_{ij}^2 x_{kl}^2}{x_{ik}^2 x_{jl}^2} \,,
\label{genu}
\ee
because these variables are invariant under the
inversion~(\ref{dualinversion}).  However, the on-shell constraints
$x_{i,i+1}^2=k_i^2=0$ imply that there are no non-trivial cross ratios
for four- and five-particle scattering.  At the six-point level there
are three non-trivial cross ratios, utilizing the allowed
$x_{i,i+2}^2$ and $x_{i,i+3}^2$ kinematic invariants:
\be
\label{defu}
\hskip-1.8cm
u_{1} = \frac{x_{13}^2 x_{46}^2}{x_{14}^2 x_{36}^2}
 = \frac{s_{12} s_{45}}{s_{123} s_{345}}\,, \quad
u_{2} = \frac{x_{24}^2 x_{51}^2}{x_{25}^2 x_{41}^2}
 = \frac{s_{23} s_{56}}{s_{234} s_{123}}\,, \quad
u_{3} = \frac{x_{35}^2 x_{62}^2}{x_{36}^2 x_{52}^2}
 = \frac{s_{34} s_{61}}{s_{345} s_{234}}\,.
\ee
As mentioned earlier, the remainder function $R_6^{(2)}$ characterizing the
two-loop MHV six-point amplitude in planar $\NeqFour$ sYM
is a function of these three variables, and this function is now known
analytically in terms of polylogarithms~\cite{DDS,Goncharov}.

In planar $\NeqFour$ sYM, dual conformal invariance combines
with dual $\NeqFour$ supersymmetry to generate dual superconformal
symmetry~\cite{DHKSdualsuper,BMdualsuper,BRTWdualsuper}.
It is natural to consider super-amplitudes,
which package sets of amplitudes together by using an $\NeqFour$ on-shell
superfield~\cite{Nair},
\be
\label{super-wave}
\hskip-2.3cm
\Phi(\eta)
 = g^{+} + \eta^A \,{\tilde g}_A
  + \frac{1}{2}\eta^A \eta^B\, \phi_{AB}
  + \frac{1}{3!}\eta^A\eta^B\eta^C \epsilon_{ABCD} \,
  {\bar{\tilde g}}^{D}
  + \frac{1}{4!}\eta^A\eta^B\eta^C \eta^D \epsilon_{ABCD} \,
  g^{-}.
\ee
Here $g^{\pm}$ are the $\pm 1$ helicity gluons, ${\tilde g}_{A}$ and
$\bar{\tilde g}^{A}$ the four flavors of $\pm \frac12$ helicity gluinos,
and $\phi_{AB}$ the six real $0$ helicity scalar states.
Integrations over the Grassmann variables $\eta^A$, $A=1,2,3,4$, can be used
to pick off the desired component amplitudes from the super-amplitudes,
defined by
\be\label{super-amplitude}
{\cal A}_n(\eta_i)
\ \equiv\ {\cal A}\Bigl( \Phi_1(\eta_1), \ldots, \Phi_n(\eta_n) \Bigr)\,.
\ee
Just as the $x_i$ bosonic variables automatically satisfy momentum
conservation, $k \equiv \sum_{i=1}^n k_i = 0$, 
one can construct their superpartners $\theta_i^{A\alpha}$,
whose differences satisfy
\be
\theta_i^{A\alpha} - \theta_{i+1}^{A\alpha}\ =\ \lambda_i^\alpha \eta_i^A \,,
\label{dualsuper}
\ee
for $i=1,2,\ldots,n$.  Super-momentum conservation constitutes eight
constraints:
\be
q^{A\alpha}\ \equiv\ \sum_{i=1}^n \lambda_i^\alpha \eta_i^A\ =\ 0.
\label{supercons}
\ee
Like ordinary momentum conservation, it is satisfied by virtue of the
periodicity of the dual coordinates, $x_i \equiv x_{i+n}$,
$\theta_i^{A\alpha} \equiv \theta_{i+n}^{A\alpha}$.  The MHV
super-amplitude generalizing \eqn{MHVPT} is
\be
{\cal A}_n^{\tree,{\rm MHV}} =
i \frac{\delta^4(k) \, \delta^8(q)}{\spa1.2\spa2.3\cdots\spa{n}.1}
= i \frac{\delta^4(x_1-x_{n+1}) \, \delta^8(\theta_1-\theta_{n+1})}
  {\spa1.2\spa2.3\cdots\spa{n}.1} \,.
\label{MHVsuper}
\ee 
It transforms covariantly under dual superconformal
transformations~\cite{DHKSdualsuper,DrummondHere} which extend SO(4,2)
to the superalgebra PSU(2,2$|$4).
Furthermore, the solutions to the BCFW super-recursion relation for amplitudes
with more negative helicities (NMHV, NNMHV, {\it etc.}) are given by
the product of this super-amplitude with collections of 
dual superconformal invariants, generically denoted by
$R_n$~\cite{DHKSdualsuper,DrummondHenn,DrummondHere}.

Superconformal and dual superconformal symmetry are separate symmetries,
which partly overlap.
They close into a very large symmetry group called the
Yangian~\cite{DHPYangian}, as discussed in this issue by Bargheer, Beisert and
Loebbert~\cite{BBLHere} and by Drummond~\cite{DrummondHere}.
The generic generator of the Yangian is not local in the sense of being a
sum of single-particle terms (like $k$ and $q$); rather it is a sum of
multi-particle operators.  (For the dual superconformal generators,
the nonlocality is mild; only the sum of two-particle operators appears.)
There are certain ``anomalies'' in the action
of the Yangian on amplitudes.  At tree level, the anomalies only act at
the boundary of the $n$-particle phase space, where two particles become
collinear, as in \fig{CollLimitFigure}~\cite{DeformSymTree}.
At loop-level they act in the full phase-space,
due to singularities in the loop integration~\cite{DeformSymLoop}.
However, it is possible to deform either the symmetry algebra
or the representation in a suitable way so as to preserve the
invariance~\cite{DeformSymTree,DeformSymLoop,BBLHere}.
It will be very interesting to see whether the constraints from the Yangian
are sufficient to determine multi-loop amplitudes directly, without having
to pass through the computation of loop integrals.

Another very recent development, in the same vein of using symmetries
or dynamical principles to bypass standard loop (or Wilson line)
integrals, concerns polygonal Wilson loops.  One considers
an operator-product-like expansion for the Wilson loop, associated
with a limit in which multiple momenta become
collinear~\cite{AGMSVOPE}.  This expansion can be carried out at both
weak and strong coupling.  At weak coupling it has been used to
compute the discontinuity of the integrated two-loop amplitude,
directly in terms of one-loop quantities~\cite{GMSVStraps}.


\section{Outlook}
\label{OutlookSection}

There has been a remarkable array of breakthroughs in our
understanding of scattering amplitudes over the past few years, as
described in the review articles in this special issue of Journal of
Physics A.  This article has represented an overview of many of these
developments, as well as providing some basic material as an
introduction to the remaining contributions.  However, it certainly
did not do justice to all of the directions being pursued currently.

One of the other remarkable features of this field has been the
interplay and cross-fertilization between formal developments and
phenomenology.  The Parke-Taylor tree amplitudes were found in the
course of trying to understand patterns arising in the structure of
five- and six-gluon tree-level amplitudes.  A similar analysis of the
structure of the five-gluon amplitude at one loop in QCD led to the
development of the unitarity method, which has been indispensable for
computing high-loop-order amplitudes in maximally supersymmetric gauge
theory and gravity, but which has also produced new NLO QCD results
for colliders~\cite{ItaHere}.  A ``radiation zero'' found over thirty
years ago in the electroweak process $d\bar{u}\to
W^-\gamma$~\cite{Mikaelian1979nr} was studied for a general gauge
theory soon thereafter~\cite{FourPtBCJ}; the relations found there were
recognized much later as the four-point versions of a more general
color-kinematics duality~\cite{BCJ08}.  Wilson loops, used to
characterize the effects of soft gluons in QCD, also turned out to
be in exact correspondence to full MHV amplitudes in planar $\NeqFour$
super-Yang-Mills theory.

This special issue should be of interest, not only to experienced
practitioners in the field, but also to newcomers who want to get
started.  As the history of the field shows, new developments can and will
come from totally unanticipated angles.
Many of the new ideas in the future may
well come to physicists whose interest in the remarkable aspects of
scattering amplitudes was sparked, at least in part, by reading the
articles assembled in this special issue.


{\bf Acknowledgments}

I am grateful to Radu Roiban, Mark Spradlin and Anastasia Volovich for
inviting me to contribute this introductory article, and I thank Radu
Roiban also for valuable comments on the manuscript.  This work
was supported by the US Department of Energy under contract
DE--AC02--76SF00515.  The figures were generated using
Jaxodraw~\cite{Jaxo1and2}, based on Axodraw~\cite{Axo}.

\clearpage



\section*{References}
\begin{thebibliography}{10}

\bibitem{Chandrasekhar}
S.~Chandrasekhar, Physics Today, {\bf 32}, issue 7, p. 25 (1979).

\bibitem{Maldacena}
J.~M.~Maldacena,
Adv.\ Theor.\ Math.\ Phys.\ {\bf 2}, 231 (1998)
[Int.\ J.\ Theor.\ Phys.\ {\bf 38}, 1113 (1999)]
[hep-th/9711200];\\
%
S.~S.~Gubser, I.~R.~Klebanov and A.~M.~Polyakov,
Phys.\ Lett.\ B {\bf 428}, 105 (1998)
[hep-th/9802109];\\
%
O.~Aharony, S.~S.~Gubser, J.~M.~Maldacena, H.~Ooguri and Y.~Oz,
Phys.\ Rept.\  {\bf 323}, 183 (2000)
[hep-th/9905111].

\bibitem{AldayMaldacena}
L.~F.~Alday and J.~M.~Maldacena,
JHEP {\bf 0706}, 064 (2007)
[0705.0303 [hep-th]].

\bibitem{ParkeTaylor}
S.~J.~Parke and T.~R.~Taylor,
Phys.\ Rev.\ Lett.\  {\bf 56}, 2459 (1986).

\bibitem{Nair}
V.~P.~Nair,
Phys.\ Lett.\  B {\bf 214}, 215 (1988).

\bibitem{ManganoParke}
M.~L.~Mangano and S.~J.~Parke,
Phys.\ Rept.\  {\bf 200}, 301 (1991)
[hep-th/0509223].

\bibitem{SWI}
M.~T.~Grisaru, H.~N.~Pendleton and P.~van Nieuwenhuizen,
Phys.\ Rev.\  D {\bf 15}, 996 (1977).
%
M.~T.~Grisaru and H.~N.~Pendleton,
Nucl.\ Phys.\  B {\bf 124}, 81 (1977).

\bibitem{SWIQCD}
S.~J.~Parke and T.~R.~Taylor,
Phys.\ Lett.\  B {\bf 157}, 81 (1985)
[Erratum-ibid.\  {\bf 174B}, 465 (1986)];\\
Z.~Kunszt,
Nucl.\ Phys.\  B {\bf 271}, 333 (1986).

\bibitem{EFK}
H.~Elvang, D.~Z.~Freedman and M.~Kiermaier,
JHEP {\bf 1010}, 103 (2010)
[0911.3169 [hep-th]].

\bibitem{EFKHere}
H.~Elvang, D.~Z.~Freedman and M.~Kiermaier,
to appear in J.\ Phys.\ A
[1012.3401 [hep-th]].

\bibitem{WittenTwistor}
E.~Witten,
Commun.\ Math.\ Phys.\  {\bf 252}, 189 (2004)
[hep-th/0312171].

\bibitem{Penrose}
R.~Penrose,
J.\ Math.\ Phys.\  {\bf 8}, 345 (1967).

\bibitem{CSW}
F.~Cachazo, P.~Svr\v{c}ek and E.~Witten,
JHEP 0409:006 (2004)
[hep-th/0403047].

\bibitem{BSTHere}
A.~Brandhuber, B.~Spence and G.~Travaglini,
to appear in J.\ Phys.\ A
[1103.3477 [hep-th]].

\bibitem{ABMSHere}
T.~Adamo, M.~Bullimore, L.~Mason and D.~Skinner,
to appear in J.\ Phys.\ A
[1104.2890 [hep-th]].

\bibitem{BCFW}
R.~Britto, F.~Cachazo, B.~Feng and E.~Witten,
Phys.\ Rev.\ Lett.\  {\bf 94}, 181602 (2005)
[hep-th/0501052].

\bibitem{BCFTree}
R.~Britto, F.~Cachazo and B.~Feng,
Nucl.\ Phys.\  B {\bf 715}, 499 (2005)
[hep-th/0412308].

\bibitem{AHCKGravity}
N.~Arkani-Hamed, F.~Cachazo and J.~Kaplan,
JHEP {\bf 1009}, 016 (2010)
[0808.1446 [hep-th]].

\bibitem{BHT2008}
A.~Brandhuber, P.~Heslop and G.~Travaglini,
Phys.\ Rev.\  D {\bf 78}, 125005 (2008)
[0807.4097 [hep-th]].

\bibitem{DrummondHenn}
J.~M.~Drummond and J.~M.~Henn,
JHEP {\bf 0904}, 018 (2009)
[0808.2475 [hep-th]].

\bibitem{DrummondHere}
J.~Drummond,
to appear in J.\ Phys.\ A.

\bibitem{BDDKNeq4}
Z.~Bern, L.~J.~Dixon, D.~C.~Dunbar and D.~A.~Kosower,
Nucl.\ Phys.\  B {\bf 425}, 217 (1994)
[hep-ph/9403226].

\bibitem{BDDKNeq1}
Z.~Bern, L.~J.~Dixon, D.~C.~Dunbar and D.~A.~Kosower,
Nucl.\ Phys.\  B {\bf 435}, 59 (1995)
[hep-ph/9409265].

\bibitem{BrittoHere}
R.~Britto,
to appear in J.\ Phys.\ A
[1012.4493 [hep-th]].

\bibitem{ItaHere}
H.~Ita,
to appear in J.\ Phys.\ A.

\bibitem{Z4partons}
Z.~Bern, L.~J.~Dixon and D.~A.~Kosower,
Nucl.\ Phys.\  B {\bf 513}, 3 (1998)
[hep-ph/9708239].

\bibitem{BCFUnitarity}
R.~Britto, F.~Cachazo and B.~Feng,
Nucl.\ Phys.\  B {\bf 725}, 275 (2005)
[hep-th/0412103].

\bibitem{BjerrumVanhove}
N.~E.~J.~Bjerrum-Bohr and P.~Vanhove,
JHEP {\bf 0804}, 065 (2008)
[0802.0868 [hep-th]];
%
JHEP {\bf 0810}, 006 (2008)
[0805.3682 [hep-th]];
%
Fortsch.\ Phys.\  {\bf 56}, 824 (2008)
[0806.1726 [hep-th]].

\bibitem{TwoLoop4g}
Z.~Bern, L.~J.~Dixon and D.~A.~Kosower,
JHEP {\bf 0001}, 027 (2000)
[hep-ph/0001001];\\
Z.~Bern, A.~De Freitas and L.~J.~Dixon,
JHEP {\bf 0203}, 018 (2002)
[hep-ph/0201161].

\bibitem{BernHuangHere}
Z.~Bern and Y.-t.~Huang,
to appear in J.\ Phys.\ A
[1103.1869 [hep-th]].

\bibitem{CJHere}
J.~J.~M.~Carrasco and H.~Johansson,
to appear in J.\ Phys.\ A
[1103.3298 [hep-th]].

\bibitem{OPP}
G.~Ossola, C.~G.~Papadopoulos and R.~Pittau,
Nucl.\ Phys.\  B {\bf 763}, 147 (2007)
[hep-ph/0609007];
JHEP {\bf 0803}, 042 (2008)
[0711.3596 [hep-ph]].

\bibitem{EGK07}
R.~K.~Ellis, W.~T.~Giele and Z.~Kunszt,
JHEP {\bf 0803}, 003 (2008)
[0708.2398 [hep-ph]].

\bibitem{BlackHat08}
C.~F.~Berger {\it et al.},
Phys.\ Rev.\  D {\bf 78}, 036003 (2008)
[0803.4180 [hep-ph]].

\bibitem{DimRed}
W.~Siegel,
Phys.\ Lett.\  B {\bf 84}, 193 (1979);\\
D.~M.~Capper, D.~R.~T.~Jones and P.~van Nieuwenhuizen,
Nucl.\ Phys.\  B {\bf 167}, 479 (1980);\\
L.~V.~Avdeev and A.~A.~Vladimirov,
Nucl.\ Phys.\  B {\bf 219}, 262 (1983).

\bibitem{FDH}
Z.~Bern and D.~A.~Kosower,
Nucl.\ Phys.\ B {\bf 379}, 451 (1992);\\
%
Z.~Bern, A.~De Freitas, L.~Dixon and H.~L.~Wong,
Phys.\ Rev.\ D {\bf 66}, 085002 (2002)
[hep-ph/0202271].

\bibitem{Sudakov}
R.~Akhoury,
Phys.\ Rev.\ D {\bf 19}, 1250 (1979);\\
A.~H.~Mueller,
Phys.\ Rev.\ D {\bf 20}, 2037 (1979);\\
%
J.~C.~Collins,
Phys.\ Rev.\ D {\bf 22}, 1478 (1980);
%
in {\it Perturbative QCD}, ed. 
A.~H. Mueller, Advanced Series on Directions in High Energy 
Physics, Vol. 5 (World Scientific, Singapore, 1989)
[hep-ph/0312336];\\
%
A.~Sen,
Phys.\ Rev.\ D {\bf 24}, 3281 (1981);
Phys.\ Rev.\ D {\bf 28}, 860 (1983).

\bibitem{MagneaSterman}
L.~Magnea and G.~Sterman,
Phys.\ Rev.\ D {\bf 42}, 4222 (1990).

\bibitem{CataniIR}
S.~Catani,
Phys.\ Lett.\ B {\bf 427}, 161 (1998)
[hep-ph/9802439].

\bibitem{TYS}
G.~Sterman and M.~E.~Tejeda-Yeomans,
Phys.\ Lett.\ B {\bf 552}, 48 (2003)
[hep-ph/0210130].

\bibitem{KM}
G.~P.~Korchemsky,
Mod.\ Phys.\ Lett.\ A {\bf 4}, 1257 (1989);\\
G.~P.~Korchemsky and G.~Marchesini,
Nucl.\ Phys.\ B {\bf 406}, 225 (1993)
[hep-ph/9210281].

\bibitem{BES}
N.~Beisert, B.~Eden and M.~Staudacher,
J.\ Stat.\ Mech.\  {\bf 0701}, P01021 (2007)
[hep-th/0610251].

\bibitem{BDS}
Z.~Bern, L.~J.~Dixon and V.~A.~Smirnov,
Phys.\ Rev.\  D {\bf 72}, 085001 (2005)
[hep-th/0505205].

\bibitem{BLSV}
J.~Bartels, L.~N.~Lipatov and A.~Sabio~Vera,
Phys.\ Rev.\  D {\bf 80}, 045002 (2009)
[0802.2065 [hep-th]];
Eur.\ Phys.\ J.\  C {\bf 65}, 587 (2010)
[0807.0894 [hep-th]].

\bibitem{AM2}
L.~F.~Alday and J.~Maldacena,
JHEP {\bf 0711}, 068 (2007)
[0710.1060 [hep-th]].

\bibitem{HexagonWilson}
J.~M.~Drummond, J.~Henn, G.~P.~Korchemsky and E.~Sokatchev,
Phys.\ Lett.\ B {\bf 662}, 456 (2008)
[0712.4138 [hep-th]].

\bibitem{BLPHere}
J.~Bartels, L.~N.~Lipatov and A.~Prygarin,
to appear in J.\ Phys.\ A
[1104.0816 [hep-th]].

\bibitem{AHPS}
L.~F.~Alday, J.~M.~Henn, J.~Plefka and T.~Schuster,
JHEP {\bf 1001}, 077 (2010)
[0908.0684 [hep-th]].

\bibitem{HennHere}
J.~M.~Henn,
to appear in J.\ Phys.\ A
[1103.1016 [hep-th]].

\bibitem{EGKMMass}
R.~K.~Ellis, W.~T.~Giele, Z.~Kunszt and K.~Melnikov,
Nucl.\ Phys.\  B {\bf 822}, 270 (2009)
[0806.3467 [hep-ph]].

\bibitem{LalRaju}
S.~Lal and S.~Raju,
Phys.\ Rev.\  D {\bf 81}, 105002 (2010)
[0910.0930 [hep-th]].

\bibitem{KLOV}
A.~V.~Kotikov, L.~N.~Lipatov, A.~I.~Onishchenko and V.~N.~Velizhanin,
Phys.\ Lett.\ B {\bf 595}, 521 (2004)
[Erratum-ibid.\ B {\bf 632}, 754 (2006)]
[hep-th/0404092].

\bibitem{Neq44}
Z.~Bern, M.~Czakon, L.~J.~Dixon, D.~A.~Kosower and V.~A.~Smirnov,
Phys.\ Rev.\  D {\bf 75}, 085010 (2007)
[hep-th/0610248].

\bibitem{NNSSublNeq4}
S.~G.~Naculich, H.~Nastase and H.~J.~Schnitzer,
JHEP {\bf 0811}, 018 (2008)
[0809.0376 [hep-th]].

\bibitem{NimaAllLoop}
N.~Arkani-Hamed, J.~L.~Bourjaily, F.~Cachazo, S.~Caron-Huot and J.~Trnka,
JHEP {\bf 1101}, 041 (2011)
[1008.2958 [hep-th]].

\bibitem{BoelsAllLoop}
R.~H.~Boels,
JHEP {\bf 1011}, 113 (2010)
[1008.3101 [hep-th]].

\bibitem{DHKSWLAnomaly}
J.~M.~Drummond, J.~Henn, G.~P.~Korchemsky and E.~Sokatchev,
Nucl.\ Phys.\  B {\bf 826}, 337 (2010)
[0712.1223 [hep-th]].

\bibitem{ABDK}
C.~Anastasiou, Z.~Bern, L.~J.~Dixon and D.~A.~Kosower,
Phys.\ Rev.\ Lett.\  {\bf 91}, 251602 (2003)
[hep-th/0309040].

\bibitem{SixPtFailure}
J.~M.~Drummond, J.~Henn, G.~P.~Korchemsky and E.~Sokatchev,
Nucl.\ Phys.\  B {\bf 815}, 142 (2009)
[0803.1466 [hep-th]];\\
Z.~Bern, L.~J.~Dixon, D.~A.~Kosower, R.~Roiban, M.~Spradlin, C.~Vergu
and A.~Volovich,
Phys.\ Rev.\  D {\bf 78}, 045007 (2008)
[0803.1465 [hep-th]].

\bibitem{BPR}
I.~Bena, J.~Polchinski and R.~Roiban,
Phys.\ Rev.\  D {\bf 69}, 046002 (2004)
[hep-th/0305116].

\bibitem{AGMTBA}
L.~F.~Alday, D.~Gaiotto and J.~Maldacena,
0911.4708 [hep-th];\\
L.~F.~Alday, J.~Maldacena, A.~Sever and P.~Vieira,
J.\ Phys.\ A  {\bf 43}, 485401 (2010)
[1002.2459 [hep-th]].

\bibitem{MHVWilsonLoops}
J.~M.~Drummond, G.~P.~Korchemsky and E.~Sokatchev,
Nucl.\ Phys.\  B {\bf 795}, 385 (2008)
[0707.0243 [hep-th]];\\
A.~Brandhuber, P.~Heslop and G.~Travaglini,
Nucl.\ Phys.\  B {\bf 794}, 231 (2008)
[0707.1153 [hep-th]];\\
J.~M.~Drummond, J.~Henn, G.~P.~Korchemsky and E.~Sokatchev,
Nucl.\ Phys.\  B {\bf 795}, 52 (2008)
[0709.2368 [hep-th]].

\bibitem{RecentWilsonLoops}
L.~F.~Alday, B.~Eden, G.~P.~Korchemsky, J.~Maldacena and E.~Sokatchev,
1007.3243 [hep-th];\\
B.~Eden, G.~P.~Korchemsky and E.~Sokatchev,
1007.3246 [hep-th];\\
L.~J.~Mason and D.~Skinner,
JHEP {\bf 1012}, 018 (2010)
[1009.2225 [hep-th]];\\
S.~Caron-Huot,
1010.1167 [hep-th];\\
A.~V.~Belitsky, G.~P.~Korchemsky and E.~Sokatchev,
1103.3008 [hep-th].

\bibitem{DDS}
V.~Del Duca, C.~Duhr and V.~A.~Smirnov,
JHEP {\bf 1003}, 099 (2010)
[0911.5332 [hep-ph]];
JHEP {\bf 1005}, 084 (2010)
[1003.1702 [hep-th]].

\bibitem{Goncharov}
A.~B.~Goncharov, M.~Spradlin, C.~Vergu and A.~Volovich,
Phys.\ Rev.\ Lett.\  {\bf 105}, 151605 (2010)
[1006.5703 [hep-th]].

\bibitem{EightOrMore}
V.~Del Duca, C.~Duhr and V.~A.~Smirnov,
JHEP {\bf 1009}, 015 (2010)
[1006.4127 [hep-th]];\\
P.~Heslop and V.~V.~Khoze,
JHEP {\bf 1011}, 035 (2010)
[1007.1805 [hep-th]].

\bibitem{KLT}
H.~Kawai, D.~C.~Lewellen and S.-H.~H.~Tye,
Nucl.\ Phys.\  B {\bf 269}, 1 (1986).

\bibitem{BCJ08}
Z.~Bern, J.~J.~M.~Carrasco and H.~Johansson,
Phys.\ Rev.\  D {\bf 78}, 085011 (2008)
[0805.3993 [hep-ph]].

\bibitem{BCJ10}
Z.~Bern, J.~J.~M.~Carrasco and H.~Johansson,
Phys.\ Rev.\ Lett.\  {\bf 105}, 061602 (2010)
[1004.0476 [hep-th]].

\bibitem{LDTASI}
L.~J.~Dixon,
in {\it QCD \& Beyond: Proceedings of TASI '95},
ed. D.\ E.\ Soper (World Scientific, 1996)
[hep-ph/9601359].

\bibitem{Cvitanovic}
P.~Cvitanovi\'{c},
{\it Group Theory}
(Nordita, 1984).

\bibitem{TreeColor}
F.~A.~Berends and W.~Giele,
Nucl.\ Phys.\  B {\bf 294}, 700 (1987);\\
M.~L.~Mangano, S.~J.~Parke and Z.~Xu,
Nucl.\ Phys.\  B {\bf 298}, 653 (1988);\\
M.~L.~Mangano,
Nucl.\ Phys.\  B {\bf 309}, 461 (1988).

\bibitem{tHooftColor}
G.~'t Hooft,
Nucl.\ Phys.\  B {\bf 72}, 461 (1974);
Nucl.\ Phys.\  B {\bf 75}, 461 (1974).

\bibitem{BKLoopColor}
Z.~Bern and D.~A.~Kosower,
Nucl.\ Phys.\  B {\bf 362}, 389 (1991).

\bibitem{BDK2f3g}
Z.~Bern, L.~J.~Dixon and D.~A.~Kosower,
Nucl.\ Phys.\  B {\bf 437}, 259 (1995)
[hep-ph/9409393].

\bibitem{DFMColor}
V.~Del Duca, A.~Frizzo and F.~Maltoni,
Nucl.\ Phys.\  B {\bf 568}, 211 (2000)
[hep-ph/9909464].

\bibitem{DDMColor}
V.~Del Duca, L.~J.~Dixon and F.~Maltoni,
Nucl.\ Phys.\  B {\bf 571}, 51 (2000)
[hep-ph/9910563].

\bibitem{KK}
R.~Kleiss and H.~Kuijf,
Nucl.\ Phys.\  B {\bf 312}, 616 (1989).


\bibitem{BCJStringProof}
N.~E.~J.~Bjerrum-Bohr, P.~H.~Damgaard and P.~Vanhove,
Phys.\ Rev.\ Lett.\  {\bf 103}, 161602 (2009)
[0907.1425 [hep-th]];\\
S.~Stieberger,
0907.2211 [hep-th].

\bibitem{Plahte}
E.~Plahte,
Nuovo Cim.\  A {\bf 66}, 713 (1970).

\bibitem{BCJRecursiveProof}
B.~Feng, R.~Huang and Y.~Jia,
Phys.\ Lett.\  B {\bf 695}, 350 (2011)
[1004.3417 [hep-th]];\\
Y.~X.~Chen, Y.~J.~Du and B.~Feng,
JHEP {\bf 1102}, 112 (2011)
[1101.0009 [hep-th]].

\bibitem{TyeZhang}
S.-H.~H. Tye and Y.~Zhang,
JHEP {\bf 1006}, 071 (2010)
[1003.1732 [hep-th]].

\bibitem{SpinorHelicity}
F.~A.~Berends, R.~Kleiss, P.~De Causmaecker, R.~Gastmans and T.~T.~Wu,
Phys.\ Lett.\  B {\bf 103}, 124 (1981);\\
P.~De Causmaecker, R.~Gastmans, W.~Troost and T.~T.~Wu,
Nucl.\ Phys.\  B {\bf 206}, 53 (1982);\\
R.~Kleiss and W.~J.~Stirling,
Nucl.\ Phys.\  B {\bf 262}, 235 (1985);\\
J.~F.~Gunion and Z.~Kunszt,
Phys.\ Lett.\  B {\bf 161}, 333 (1985);\\
Z.~Xu, D.~H.~Zhang and L.~Chang,
Nucl.\ Phys.\  B {\bf 291}, 392 (1987);\\
R.~Gastmans and T.~T.~Wu,
{\it The Ubiquitous Photon: Helicity Method for QED and QCD} 
(Clarendon Press, 1990).

\bibitem{SchabingerHere}
R.~Schabinger,
to appear in J.\ Phys.\ A
[1104.3873 [hep-th]].

\bibitem{DHKSdualsuper}
J.~M.~Drummond, J.~Henn, G.~P.~Korchemsky, E.~Sokatchev,
Nucl.\ Phys.\  B {\bf 828}, 317 (2010).
[0807.1095 [hep-th]].

\bibitem{BMdualsuper}
N.~Berkovits and J.~Maldacena,
JHEP {\bf 0809}, 062 (2008)
[0807.3196 [hep-th]].

\bibitem{BRTWdualsuper}
N.~Beisert, R.~Ricci, A.~A.~Tseytlin and M.~Wolf,
Phys.\ Rev.\  D {\bf 78}, 126004 (2008)
[0807.3228 [hep-th]].

\bibitem{DHPYangian}
J.~M.~Drummond, J.~M.~Henn and J.~Plefka,
JHEP {\bf 0905}, 046 (2009)
[0902.2987 [hep-th]].

\bibitem{BBLHere}
T.~Bargheer, N.~Beisert and F.~Loebbert,
to appear in J.\ Phys.\ A
[1104.0700 [hep-th]].

\bibitem{DeformSymTree}
T.~Bargheer, N.~Beisert, W.~Galleas, F.~Loebbert and T.~McLoughlin,
JHEP {\bf 0911}, 056 (2009)
[0905.3738 [hep-th]].

\bibitem{DeformSymLoop}
N.~Beisert, J.~Henn, T.~McLoughlin and J.~Plefka,
JHEP {\bf 1004}, 085 (2010)
[1002.1733 [hep-th]].

\bibitem{AGMSVOPE}
L.~F.~Alday, D.~Gaiotto, J.~Maldacena, A.~Sever and P.~Vieira,
JHEP {\bf 1104}, 088 (2011)
[1006.2788 [hep-th]].

\bibitem{GMSVStraps}
D.~Gaiotto, J.~Maldacena, A.~Sever and P.~Vieira,
1102.0062 [hep-th].

\bibitem{Mikaelian1979nr}
K.~O.~Mikaelian, M.~A.~Samuel and D.~Sahdev,
Phys.\ Rev.\ Lett.\  {\bf 43}, 746 (1979).

\bibitem{FourPtBCJ}
D.~Zhu,
Phys.\ Rev.\  D {\bf 22}, 2266 (1980);\\
C.~J.~Goebel, F.~Halzen and J.~P.~Leveille,
Phys.\ Rev.\  D {\bf 23}, 2682 (1981).

\bibitem{Jaxo1and2}
D.~Binosi and L.~Theussl,
Comput.\ Phys.\ Commun.\ {\bf 161}, 76 (2004)
[hep-ph/0309015];\\
D.~Binosi, J.~Collins, C.~Kaufhold and L.~Theussl,
Comput.\ Phys.\ Commun.\  {\bf 180}, 1709 (2009)
[0811.4113 [hep-ph]].

\bibitem{Axo}
J.~A.~M.~Vermaseren,
Comput.\ Phys.\ Commun.\ {\bf 83}, 45 (1994).

\end{thebibliography}
\end{document}